\documentclass[twocolumn,secnumarabic,amssymb, nobibnotes, aps, prd]{revtex4}
\usepackage{graphicx}

\begin{document}
\title{Magnetic dependence of the critical and persistent current of asymmetric superconducting rings}
\author{V. L. Gurtovoi, S. V. Dubonos, A. V. Nikulov, N.N.Osipov, and V. A. Tulin}
\affiliation{Institute of Microelectronics Technology and High Purity Materials, Russian Academy of Sciences, 142432 Chernogolovka, Moscow District, RUSSIA. } 
\begin{abstract} The obtained periodic magnetic-field dependencies of the critical current $I_{c+}(\Phi /\Phi _{0})$, $I_{c-}(\Phi /\Phi _{0})$, measured in opposite directions on asymmetric superconducting aluminum rings, allow to explain observed earlier quantum oscillations of a dc voltage as a result of alternating current rectification. It is found, that the high efficiency of the rectification of both individual rings and ring systems is connected to a hysteresis of the current-voltage characteristics. The asymmetry of the current-voltage characteristics providing the rectification effect is due to the relative shifts of the magnetic dependencies $I_{c-}(\Phi /\Phi _{0}) = I_{c+}(\Phi /\Phi _{0}+\Delta \phi )$ of the critical current measured in opposite directions. This shift means  that position of $I_{c+}(\Phi /\Phi _{0})$ and $I_{c-}(\Phi /\Phi _{0})$ minimums does not correspond to  $n + 0.5$ magnetic flux quantum $\Phi _{0}$ which is in the direct contradiction with measurement results of the Little-Parks resistance oscillations. Despite of this contradiction the amplitude of the critical current anisotropy oscillations $I_{c,an}(\Phi /\Phi _{0}) = I_{c+}(\Phi /\Phi _{0}) - I_{c-}(\Phi /\Phi _{0})$ and its variations with temperature correspond to expected amplitude of the persistent current oscillations and to its variations with temperature.
 \end{abstract}

\maketitle

\narrowtext

\section{INTRODUCTION}

It is known that the current $I = \oint_{l} dl E/R_{l}$  induced by the Faraday electromotive force $ \oint_{l} dl E = -d\Phi /dt$, flowing along a conducting loop $l$, should cause the potential difference $V = (R_{ls} - R_{l}l_{s}/l)I = R_{anym}I$ on the segment $l_{s}$ with the resistance $R_{ls}$ different $R_{ls}/l_{s} \neq R_{l}/l$ from the resistance $R_{l}$ along the whole circle $l$. 

Measurements have discovered that the potential difference $V_{dc}(\Phi /\Phi _{0})$ can be observed on an asymmetric superconducting loop with \cite{QuOs1967} and without Josephson junctions \cite{QuOs2001,QuOs2003}. The sign and magnitude of this potential difference periodically depend on the magnetic flux $\Phi $ inside the loop; the period is equal to the flux quantum $\Phi _{0} = \pi \hbar /e$. It is known \cite{Tinkham} that the averaged value of the persistent current $\overline{I_{p}}$ circulating in such loops depends on the magnetic flux in a similar manner $\overline{I_{p}}(\Phi /\Phi _{0})$. Thus, there is a certain analogy between the persistent current \cite{Tinkham}, i.e., the direct current existing due to Bohr quantization under equilibrium conditions at the magnetic flux $\Phi \neq n\Phi _{0}$ constant in time $d\Phi /dt = 0$, and the ordinary circulating current existing due to the Faraday electromotive force, i.e., at $d\Phi /dt \neq 0$. Since the observed dependencies $V_{dc}(\Phi /\Phi _{0})$ \cite{QuOs1967,QuOs2001,QuOs2003} are proportional to $\overline{I_{p}}(\Phi /\Phi _{0})$ we can write $V_{dc}(\Phi /\Phi _{0}) = R_{as}\overline{I_{p}}(\Phi /\Phi _{0})$ introducing the quantity $R_{as}$ by analogy with the resistance difference $(R_{ls} - R_{l}l_{s}/l)$ defining the relation between the potential difference and circulating current in an ordinary asymmetric loop. We emphasize that observation of oscillations $V_{dc}(\Phi /\Phi _{0})$ on superconducting structures both with \cite{QuOs1967} and without \cite{QuOs2001,QuOs2003} Josephson junctions confirms the analogy between a one dimensional ring and the superconducting quantum interferometer, noted in some papers \cite{Fink1987}. 

The quantity $R_{as}$ in a superconducting ring, unlike an ordinary ring, is not constant, but depends on various factors, first of all, on temperature $T$. It is clear that $R_{as} \neq 0$ and the potential difference $V_{dc}(\Phi /\Phi _{0}) = R_{as}\overline{I_{p}}(\Phi /\Phi _{0})$ can be observed only in the resistive state. The ring resistance in an equilibrium state is nonzero, $R_{l} > 0$, at temperatures higher, $T > T_{c}$, or close, $T \approx  T_{c}$, to the superconducting transition temperature $T_{c}$. Observations by Little and Parks \cite{LP1962} of resistance oscillations $R(\Phi /\Phi _{0})$ of a thin-walled superconducting cylinder in a magnetic field showed that not only $R_{l} > 0$, but also $I_{p}(\Phi /\Phi _{0}) \neq 0$ at temperatures $T \approx  T_{c}$ corresponding to the resistive transition $R_{ln} > R_{l} > 0$. According to the conventional explanation \cite{Tinkham}, the periodic variation in the resistance $R(\Phi /\Phi _{0})$ of a thin-walled superconducting cylinder \cite{LP1962} or a ring \cite{Mosh1992}, measured at a temperature corresponding to the resistive transition $R_{ln} > R_{l} > 0$, is caused by the change in the critical temperature $\Delta R(\Phi /\Phi _{0}) \propto -\Delta T_{c}(\Phi /\Phi _{0})$. The decrease in $T_{c}$ at $\Phi \neq \Phi _{0}$, $-\Delta T_{c}(\Phi /\Phi _{0}) \propto v_{s}^{2}(\Phi /\Phi _{0}) \propto (n - \Phi /\Phi _{0})^{2}$ is related to an increase in the superconducting state energy in proportion to $v_{s}^{2}(\Phi /\Phi _{0}) $, when the permitted superconducting pair velocities $v_{s}(\Phi /\Phi _{0}) \propto (n - \Phi /\Phi _{0})$ cannot be zero \cite{Tinkham,Fink1987}. Little-Parks oscillations  $\Delta R(\Phi /\Phi _{0}) = \Delta V(\Phi /\Phi _{0})/I_{ext}$ are observed at the measured current $I_{ext}$, which can differ by hundreds of times \cite{Mosh1992}; there is no reason to doubt that they should also be observed in the limit of infinitesimal values of $I_{ext}$ , i.e., at thermodynamic equilibrium. The persistent current $I_{p} = sj_{p} = s2en_{s}v_{s} \propto    n - \Phi /\Phi _{0}$ observed due to the quantization of the velocity circulation of superconducting pairs \cite{Tinkham},
$$\oint dl v = \frac{2\pi \hbar }{m}(n - \frac{\Phi}{\Phi _{0}} ) \eqno{(1)}$$
is an equilibrium phenomenon. At nonzero resistance $R_{l} > 0$, the persistent current can be observed due to thermal fluctuations \cite{Kulik70}, which switch the ring between superconducting states with different connectivity of the wave function. 

At $T < T_{c}$, where $R_{l} = 0$ in equilibrium state, these phenomena are observed at an external current exceeding the critical superconducting current $I_{c}(T)$. In the case of Little-Parks oscillation, this corresponds to the shift of the resistive transition $T_{c}(I_{ext})$ by direct measuring current $I_{ext}$ \cite{Mosh1992}. Quantum oscillations $V_{dc}(\Phi /\Phi _{0})$ of the dc voltage are induced at $T < T_{c}$ by alternating current \cite{QuOs2003} or uncontrollable electrical noise \cite{QuOs1967,QuOs2001}. The results of the measurements \cite{QuOs2003} on asymmetric aluminium rings showed that quantum oscillations $V_{dc}(\Phi /\Phi _{0})$ of the dc voltage at $T < T_{c}$ are induced by the external alternating current, e.g., $I_{ext} = I_{0}sin(2\pi ft)$, when its amplitude $I_{0}$ exceeds the critical value $I_{0c}$ close to the critical current $I_{c}(T)$. It was shown that the result is independent of the alternating current frequency $f$ in the studied frequency range $f = 100 \ Hz \div 1 \ MHz$, and the amplitude $V_{A}$ of $V_{dc}(\Phi /\Phi _{0})$ quantum oscillation non-monotonically depends on the current amplitude $I_{0}$ \cite{QuOs2003}. As the $V_{dc}(\Phi /\Phi _{0})$ amplitude reaches the maximum $V_{A} = V_{A, max}$ at $I_{0} = I_{0, max}$, it decreases as the external current amplitude $I_{0}$ further increases (see Figs. 11 and 12).

This result can be explained in two consistent ways. One explanation proceeds from the relation $V_{dc}(\Phi /\Phi _{0}) = R_{as}\overline{I_{p}}(\Phi /\Phi _{0})$ obtained above by the analogy between persistent and ordinary currents. The non-monotonic dependence of the amplitude $V_{A}$ of quantum oscillations $V_{dc}(\Phi /\Phi _{0})$ on the external current amplitude $I_{0}$ can be explained as a result of the superconducting transition shift by the current $I_{ext}$. The potential difference is nonzero, $V_{dc}(\Phi /\Phi _{0}) \neq 0$ at $R_{as} \neq 0$ and $\overline{I_{p}}(\Phi /\Phi _{0}) \neq 0$. Above the superconducting transition temperature, the resistance is nonzero, $R_{as} \neq 0$, but the persistent current is zero, $I_{p}(\Phi /\Phi _{0}) = 0$; while $I_{p}(\Phi /\Phi _{0}) \neq 0$, but $R_{as} = 0$ below the superconducting transition. At small current amplitude $I_{0}$, the shift of the superconducting transition temperature $T_{c}(I_{ext})$ does not reach the temperature $T < T_{c}$ at which measurements are performed, therefore $V_{dc}(\Phi /\Phi _{0}) = 0$. Quantum oscillations $V_{dc}(\Phi /\Phi _{0})$ become observable, when $I_{0}$ lowers the superconducting transition temperature to the measurement temperature $T_{c}(I_{0}) \approx  T < T_{c}(0)$. In this case, they initially become observable under strong magnetic fields which promote the superconducting transition shift to low temperatures (see Fig. 11). The amplitude $V_{A}$ reaches a maximum at $I_{0}$ corresponding to the current which shifts the transition by $T - T_{c}(0)$ and decreases as $I_{0}$ further increases due to the persistent current disappearance into a normal state.

The second explanation proposed in \cite{QuOs2003} interprets quantum oscillations $V_{dc}(\Phi /\Phi _{0})$ as a result of rectification of the external alternating current due to the asymmetry of current-voltage characteristic (CVCs) of the asymmetric ring. Such a ring with the geometry used in both \cite{QuOs2003} and this study is shown in Fig. 1. To explain the observed oscillations $V_{dc}(\Phi /\Phi _{0})$, the CVC asymmetry sign and magnitude should be periodic functions of the magnetic field. Such periodic variations in the asymmetry are explained in \cite{QuOs2003} as a result of the superposition of external $I_{ext}$ and persistent $I_{p}(\Phi /\Phi _{0})$ currents. According to this explanation, the persistent current in a ring with unequal cross sections of halves (see Fig.1) can be found as a quantity proportional to the difference of critical currents measured in opposite directions. 

The goal of this study is to test the model explaining quantum oscillations $V_{dc}(\Phi /\Phi _{0})$ as a result of alternating current rectification and to study in detail the correlation of the CVC behaviour and asymmetry with the rectification efficiency in structures with different geometries. The temperature dependencies of the maximum amplitude $V_{A, max}$ of quantum oscillations, the amplitude $I_{0,max}$ of the current $I_{ext}$ at which oscillations are observed, and the critical amplitude $I_{0c}$ of the $I_{ext}$ current were studied in detail. The dependencies of the persistent current as a quantity proportional to the critical current anisotropy on the magnetic field and temperature were obtained.

To solve the posed problem, first of all, CVCs of the systems under consideration were measured at various temperatures. The results of these measurements are given in Section 3.1. The magnetic dependencies $I_{c+}(\Phi /\Phi _{0})$ and $I_{c-}(\Phi /\Phi _{0})$ of the critical current measured in opposite directions of the external current $I_{ext}$ are presented in Section 3.2. In Section 3.3, it is shown that the voltage $V_{dc}(\Phi /\Phi _{0})$ is a result of alternating current rectification on asymmetric rings whose CVC asymmetry varies periodically in a magnetic field due to periodic variations of the anisotropy of critical current $I_{c+}(\Phi /\Phi _{0})$ and $I_{c-}(\Phi /\Phi _{0})$. It is emphasized that the rectification mechanism near $T_{c}$ where CVCs are reversible, i.e., no hysteresis is observed, can differ from that at lower temperatures where a significant CVC hysteresis is observed. The measured temperature dependencies of the maximum amplitude $V_{A, max}$ of quantum oscillations $V_{dc}(\Phi /\Phi _{0})$, the amplitude $I_{0, max}$ of the current inducing these oscillations, and the rectification efficiency defined as the ratio of these parameters are given in Section 3.4. The observed amplitude of oscillations of the critical current and its anisotropy are compared with the expected amplitude of persistent current oscillations in Section 3.5. The qualitative differences of observed critical current oscillations from the expected ones are considered in Section 3.6. In the Conclusions, we note that the results obtained allow us to explain the previously observed phenomena; however, pose new problems requiring further study. 

\section{EXPERIMENTAL DETAILS}
The study was conducted in the temperature range $T = 1.19 \div 1.3 \ K$ on aluminum film nanostructures with the superconducting transition temperature $T_{c} \approx  1.23 \div 1.27 \ K$, the resistance per square of $\approx 0.5 \ \Omega $  at $4.2 \ K$, and the resistance ratio $R(300 K)/R(4.2 K) \approx  3$. The diameter of all rings was $2r = 4 \ \mu m$, the widths of ring halves $l_{w}$ and $l_{n}$ were $w_{w} = 0.4 \ \mu m$ and $w_{n} = 0.2 \ \mu m$, respectively (see Fig. 1), the thickness was $d = 40 \div 50 \ nm$; the cross sections were, respectively, $s_{w} = w_{w}d \approx  0.016 \div  0.02 \ \mu m^{2}$ and $s_{n} = w_{n}d \approx  0.008 \div  0.01 \ \mu m^{2}$. The measurements were performed on four samples. The first two represented a system of 20 rings connected in series with current contacts of width  $w_{con} = 0.4 \ \mu m$ (in this geometry, the signal was picked up from 18 rings) and  $w_{con} = 1 \ \mu m$ between them. The second pair consisted of single rings; in this case, the width of current contacts was  $w_{con} = 0.6 \ \mu m$ and $w_{con} = 0.7 \ \mu m$. The microstructures were fabricated on Si substrates using a JEOL-840A scanning electron microscope transformed into a laboratory electron lithograph using the NANOMAKER software package.

\begin{figure}[b]
\includegraphics{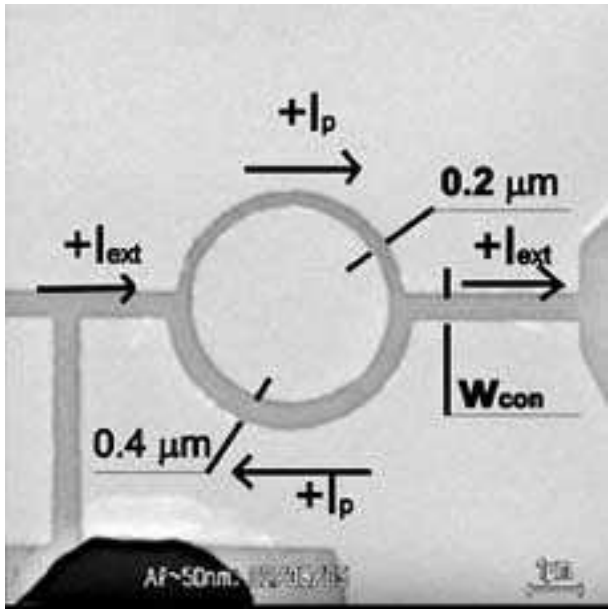}
\caption{\label{fig:epsart} Micrograph of the structure used in measurements. The width of ring halves is $w_{w} = 0.4 \ \mu m$ and $w_{s} = 0.2 \ \mu m$; the width of contacts supplying the external current is $w_{con} = 0.4 \ \mu m$.}
\end{figure}

Measurements were performed by the four-probe method in a glass helium cryostat using $He^{4}$ as a coolant whose pumping allowed a temperature decrease down to $1.19 \ K$. Direct, sawtooth, or sinusoidal current from a Keithley 6221 precision source was applied to current contacts. The direct current was used to measure the dependences of the sample resistance on the magnetic field $R(B)$ (Little-Parks oscillations) and temperatures $R(T)$. The sawtooth current was used to record CVCs and the dependencies of the critical current on the temperature $I_{c}(T)$ and magnetic field $I_{c}(B)$. The sinusoidal current, along with the sawtooth current, was used to study the rectification effect and to measure the dependencies of the rectified voltage on the magnetic field $V_{dc}(B)$. The voltage was measured on potential contacts by an instrumental amplifier with a gain of 1000 and a noise level (reduced to the input) of $20 \ nV$ in a frequency band from zero to $1 \ Hz$. Then the signal was amplified by an SR560 (Stanford Research Systems) preamplifier used for additional amplification and the formation of a required signal band by low- and highfrequency filters. The temperature was measured using a calibrated thermistor ($R(300 K) = 1.5 \ k\Omega $) with a measuring current of $0.1 \ \mu A$. The amplified voltage from the sample and the signals proportional to the sample current, magnetic field, and temperature were simultaneously digitized by a 16-bit ADC with eight differential inputs.

The magnetic field $B$ perpendicular to the sample plane was induced by a copper solenoid placed outside the cryostat. The dependences of the measured values on the solenoid current $I_{sol}$ were recorded. The magnetic field induced by the solenoid current was determined by the calibration $B_{sol} = k_{sol}I_{sol}$ performed using a Hall sensor probe, $k_{sol} = 129 \ G/A$. The period $B_{0} = \Phi _{0}/S = 1.4 \div 1.6 \ G$ of all periodic dependences $R(B)$, $V_{dc}(B)$, and $I_{c}(B)$ corresponded to the ring area $S = \pi r^{2} = 14.8 \div 13.0 \ \mu m^{2}$ used in measurements, where $r = 2.2 \div 2.0 \ \mu m$ is the quantity close to the inner radius of a given ring. To reduce the geomagnetic field, the cryostat region with a sample was shielded by a Permalloy cylinder. The residual magnetic field was $B_{res} \approx  0.15 \ G$, i.e., about one-tenth of the period $B_{0}$. Due to partial shielding, the resistance minimum $R(B_{sol})$ and the zero value of the rectified voltage $V_{dc}(B_{sol})$ were observed at $B_{sol} = -B_{res} \approx  -0.15 \ G$. In this paper, we present all the measured dependences as functions of the magnetic flux inside the ring, $\Phi  = SB = S(B_{sol} + B_{res})$, induced by external fields $B_{sol} + B_{res}$. The exact value of the area $S$ was chosen from the condition of the equality of the period of oscillations $R(\Phi )$, $V_{dc}(\Phi )$, and $I_{c}(\Phi )$ to the magnetic flux quantum $\Phi _{0}$. The exact value of the residual magnetic field $B_{res}$ was chosen from the condition of $R(\Phi )$ minimum at $\Phi  = 0$, $V_{dc}(\Phi ) = 0$ at $\Phi  = 0$, and the condition $I_{c-}(\Phi ) = I_{c+}(-\Phi )$. In all cases, the value of $B_{res}$ is about the same. When plotting the dependencies $R(\Phi )$, $V_{dc}(\Phi )$, and $I_{c}(\Phi )$, the magnetic flux $\Phi _{I} = \Phi _{Iext} + \Phi _{Ip}$ induced by the external $I_{ext}$ and persistent $I_{p}$ currents was neglected. We used the approximation $\Phi  = \Phi _{ext} + \Phi _{I} \approx  \Phi _{ext} = S(B_{sol} + B_{res})$, since the flux induced by the external $\Phi _{Iext}$ and persistent $\Phi _{Ip}$ currents does not exceed a few hundredths of the flux quantum in the studied temperature range at the inductance $L \approx  2 \ 10^{-11} \ H$ of rings used in this study and their low critical and persistent currents. This problem is considered in more detail in Section 3.6.

The dependences $R(B)$, $V_{dc}(B)$, and $I_{c}(B)$ were recorded during sawtooth temporal variations of the solenoid current $I_{sol}(t)$ with a frequency of $0.01 \div 0.1 \ Hz$. The amplitude of magnetic field variations did not exceed $100 \ G$. Little-Parks oscillations were recorded using the direct current $I_{ext} = 0.1 \div 2 \ \mu A$. The dependencies of the rectified voltage $V_{dc}(\Phi )$ were measured using the sinusoidal or sawtooth current $I_{ext}(t)$ with an amplitude up to $50 \ \mu A$ and a frequency of $0.5 \div 5 \ kHz$. The width of the rectified voltage spectrum caused by magnetic field variations did not exceed $1 \ Hz$; therefore, the rectified voltage was amplified in a frequency band from zero to $10 \div 30 \ Hz$. On the one hand, this did not result in signal shape distortions; on the other hand, this prevented ac voltage appearance at the amplifier output.

The dependences of the critical current on the magnetic field, $I_{c+}(B)$ and $I_{c-}(B)$), were measured from CVCs periodically ($10 \ Hz$) repeated in a slowly varying magnetic field ($\approx 0.01 \ Hz$) by the following algorithm: (i) the condition of the structure existence in the superconducting state was tested and then (ii), as the threshold voltage (set above interferences and noises of the measuring system and controlling the minimum measurable critical current) was exceeded, the magnetic field and critical current were determined with a delay of $\approx 30 \ \mu s$. Thus, the critical was sequentially measured in positive $I_{c+}$ and negative $I_{c-}$ directions of the measuring current $I_{ext}$. The measurement of a single dependence $I_{c+}(B)$, $I_{c-}(B)$ containing 1000 points took $\approx 100 \ s$.

\section{RESULTS AND DISCUSSION}
To test the hypothesis about observation of quantum oscillations of the direct voltage $V_{dc}(\Phi /\Phi _{0})$ as a result of alternating current rectification, CVCs and the dependence of their anisotropy on the magnetic field should be studied for the structures under consideration.

\subsection{Current-Voltage Characteristics}
Measurements showed three CVC types which should be distinguished to describe the rectification process. We attribute smooth reversible CVCs to the first type (Figs. 2-4). This type is observed for all structures near the superconducting transition: in the measurements on 18 rings with $w_{con} = 0.4 \ \mu m$ at $T > 0.993 \ T_{c}$ (Fig. 2), on 20 rings with $w_{con} = 1 \ \mu m$ at $T > 0.985 \ T_{c}$ (Fig. 3), and on single rings at $T > 0.997 \ T_{c}$ (Fig. 4). The second and third types include CVCs (measured in the current source mode) with hysteresis (Figs. 2-4). The third type is characterized by a sharp transition of the entire structure to the resistive state (Figs. 2, 4). In contrast to this, in the second-type CVC, this transition is observed in a certain range of the current $I_{ext}$ (Figs. 2, 3). The second-type CVCs were observed for the 20-ring system in an intermediate temperature range, $0.993 \ T_{c} > T > 0.985 \ T_{c}$, on a system with $w_{con} = 0.4 \ \mu m$ (Fig. 2), and in the entire range below $T < 0.985 \ T_{c}$ on a system with $w_{con} = 1 \ \mu m$ (Fig. 3). In the measurements on a single ring, no second-type CVCs were observed (Fig. 4). The third-type CVCs are observed at low temperatures, $T < 0.985 \ T_{c}$, in the measurements on 18 rings with $w_{con} = 0.4 \ \mu m$ and $T < 0.995 \ T_{c}$ on a single ring. 

The difference of the 20-ring structure with $w_{con} = 1 \ \mu m$ from the structure with $w_{con} = 0.4 \ \mu m$ is the absence of the third-type CVCs down to the lowest temperatures at which the measurements were performed (Fig. 3). In contrast to this, CVCs of single rings belong to the third type in almost the entire temperature range up to $T = 0.995 \ T_{c}$ (Fig. 4). In this case, a sharp change in the resistance is observed not only during the transition of the single ring to the resistive state, but also during the reverse transition from the resistive to the superconducting state (Fig. 4), which is a distinctive feature of single rings in comparison with a ring system (see Figs. 2 and 3). The difference is also observed in the shape of resistive transitions: the lower part of the resistive transition of the single ring is narrower than that of the ring system (see Fig. 15). We relate these qualitative differences between resistive characteristics of the single ring and a long chain of rings to the proximity effect of wide superconducting areas arranged at a distance of several $\mu m$ from the single ring. The proximity effect on resistive characteristics of aluminum structures appears in an anomalous increase in the resistance before the superconducting transition (see Fig. 15). Such an anomaly was observed by many authors \cite{Anomaly} in measurements on aluminum structures with sizes comparable to the correlation length $\xi (T)$. This anomaly, as well as others \cite{Anomaly}, has not yet been reasonably explained and requires further study. In this paper, the features of resistive characteristics of aluminum structures, including the anomaly, are considered only for better understanding of alternating current rectification resulting in $V_{dc}(\Phi /\Phi _{0})$.

The third-type CVCs (curves in Figs. 2 and 4) are typical of superconducting nanostructures with close critical currents of their elements, when the transition to the resistive state occurs due to exceeding the depairing current density \cite{Tinkham}, 
$$j_{c} = \frac{H_{c}(T)}{3\surd 6\lambda _{L}(T)} = \frac{H_{c}(0)}{3\surd 6\lambda _{L}(0)}(1 - \frac{T}{T_{c}})^{3/2}$$
For aluminum with the thermodynamic critical field extrapolated to the zero temperature, $H_{c}(0) = 105 \ Oe$, the London penetration depth $\lambda _{L}(0) = 50 \ nm$, and the depairing current density is estimated as $j_{c} = j_{c}(0)(1 - T/T_{c})^{3/2} \approx  9 \ 10^{6} \ A/cm^{2} (1 - T/T_{c})^{3/2}$. The experimental temperature dependences of the critical current in the studied range $T = 0.965 \div  0.995 \ T_{c}$ can be described as $I_{c}(T) = I_{c}(0)(1 - T/T_{c})^{3/2}$, where $I_{c}(0) \approx  1.4 \ mA$ for an 18-ring system with $w_{con} = 0.4 \ \mu m$ (the minimum cross section is $s_{con} \approx  0.016 \ \mu m^{2}$), $I_{c}(0) \approx  3 \ mA$ for the 20-ring system with $w_{con} = 1 \ \mu m$ (the minimum cross section is $s_{w} + s_{n} \approx  0.03 \ \mu m^{2}$), $I_{c}(0) \approx  5.5 \ mA$ and $I_{c}(0) \approx  4.3 \ mA$ for single rings with $w_{con} = 0.7 \ \mu m$ and $w_{con} = 0.6 \ \mu m$, respectively. The critical current density ($j_{c}(0) \approx  9 \ 10^{6} \ A/cm^{2}$ for the 18-ring system and $j_{c}(0) \approx  10^{7} \ A/cm^{2}$ for the 20-ring system) is close to theoretical values. The critical current density of single rings ($j_{c}(0) \approx  2.3 \ 10^{7} \ A/cm^{2}$ for a ring with $w_{con} = 0.7 \ \mu m$ and $j_{c}(0) \approx  1.8 \ 10^{7} \ A/cm^{2}$ for a ring with $w_{con} = 0.6 \ \mu m$) is somewhat higher. The increased critical current in single-ring structures can be caused by the effect of wide superconducting banks due to a relatively short length of the structure with a minimum cross section of $\approx 0.024 \ \mu m^{2}$ in comparison with the correlation length $\xi (T)$. 

\begin{figure}
\includegraphics{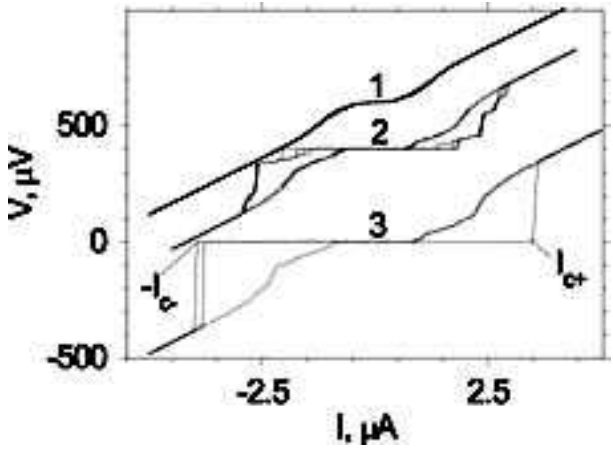}
\caption{\label{fig:epsart} CVC of the system of 18 asymmetric rings connected in series with $w_{con} = 0.4 \ \mu m$ at temperatures (1) $T \approx  1.261 \ K \approx  0.995T_{c}$, (2) $T \approx  1.252 \ K \approx  0.988T_{c}$, and (3) $T \approx  1.247 \ K \approx  0.984T_{c}$.}
\end{figure}

\begin{figure}
\includegraphics{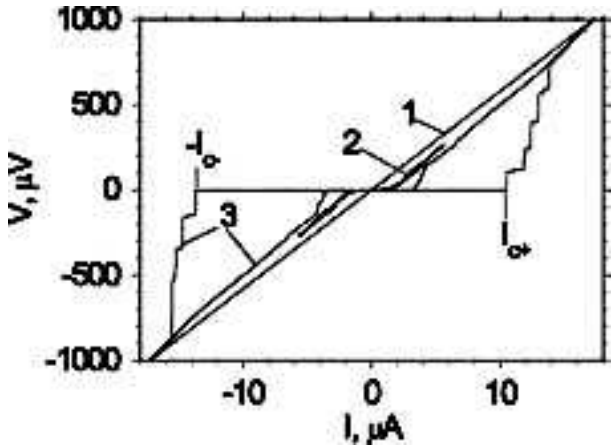}
\caption{\label{fig:epsart} CVC of the systems of 20 asymmetric rings connected in series with $w_{con} = 1 \ \mu m$ at temperatures (1) $T > T_{c}$, (2) $T \approx  1.222 \ K \approx  0.988T_{c}$, and (3) $T \approx  1.199 \ K \approx  0.969T_{c}$. }
\end{figure}

\begin{figure}
\includegraphics{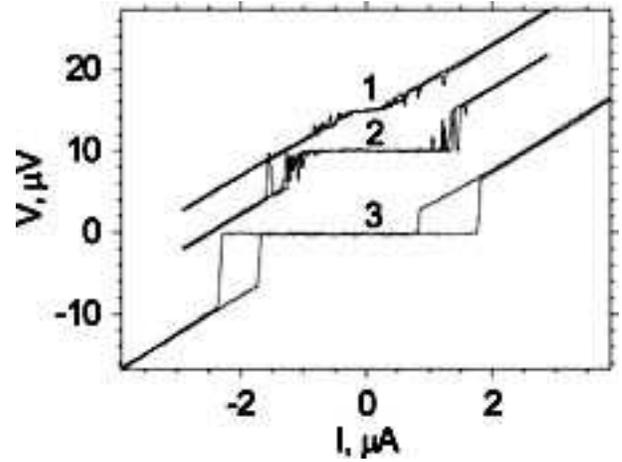}
\caption{\label{fig:epsart} CVC of the single asymmetric ring with $w_{con} = 0.7 \ \mu m$ at temperatures (1) $T \approx  1.243 \ K \approx  0.999T_{c}$, (2) $T \approx  1.241 \ K \approx  0.997T_{c}$, and (3) $T \approx  1.239 \ K \approx  0.995T_{c}$. }
\end{figure}

The sharp transition to the resistive state of the whole structure at once and the significant CVC hysteresis are caused by the nonmonotonic dependence of the superconducting current density $j_{s} = 2en_{s}v_{s}$ on the superconducting pair velocity $v_{s}$ comparable to the depairing velocity $v_{sc} = \hbar /m \surd 3\xi (T)$ \cite{Tinkham}. The transition to the normal state, $n_{s} = 0$, as the critical velocity $v_{s} = v_{sc}$ of superconducting pairs is reached by any structure region, results in a decrease in the pair density $n_{s}$ in neighboring regions at distances of the order of the correlation length $\xi (T)$, hence, in an increase in the pair velocity at a given current $I_{ext} = I_{s} = sj_{s} = s2en_{s}v_{s}$. An increase in the velocity above the critical value results in the transition of these neighboring regions to the normal state, if their cross section area slightly exceeds the least cross section area. Therefore, when the current density reaches the depairing current density in one of segments, the whole structure transits to the resistive state, even a long 18-ring system connected in series with a full length of $176 \ \mu m$ (see Fig. 2). The proximity effect on the pair velocity in neighboring regions of the structure should weaken as the cross section of isolated regions increases and the equilibrium pair density decreases. Therefore, the third-type CVCs are not observed in the 20-ring structure with $w_{con} = 1 \ \mu m$ (Fig. 3); in the 18-ring system with $w_{con} = 0.4 \ \mu m$, the transition from the third to the second CVC type as $T_{c}$ is approached is observed (Fig. 2). The hysteresis is observed because of the appearance of the nonzero density $n_{s}$ of superconducting pairs at a given current $I_{ext}$ can be stable only if $n_{s} > I_{ext}/s2ev_{sc}$. Otherwise, pairs should be accelerated in an electric field to the depairing velocity and disappear. The formation of the high pair density on sufficiently large length scales has a low probability; therefore, the return to the superconducting state occurs at lower $I_{ext}$ than the transition to the resistive state (Figs. 2-4). The CVC hysteresis is observed if a change in the superconducting pair density due to fluctuations does not exceed $n_{s} = I_{ext}/s2ev_{sc}$. This takes place at low temperatures, when the equilibrium pair density is significant and the kinetic energy of pairs,
$$E_{kin} = \frac{\lambda _{L}^{2}(T)\xi (T)}{2s}\mu _{0}I_{c}^{2}(T)$$
corresponding to the depairing current $I_{c}$, is much higher than the thermal fluctuation energy $k_{B}T$ in a volume equal to the product of the superconductor cross section $s$ and the correlation length $\xi (T)$. Near $T_{c}$, the kinetic energy corresponding to the depairing current $I_{c} \propto  (1 - T/T_{c})^{3/2}$ becomes comparable to the thermal fluctuation energy,
$$E_{kin} = \frac{\lambda _{L}^{2}(T)\xi (T)}{2s}\mu _{0}I_{c}^{2}(T) \approx k_{B}T$$
Using the values $\lambda _{L}(0) \approx  50 \ nm$, $\xi (0) = 170 \ nm$, $s = 0.016 \mu m^{2}$, $I_{c}(T) = I_{c}(0)(1 - T/T_{c})^{3/2} = 1.4 \ mA \times (1 - T/T_{c})^{3/2}$, we find that
$$E_{kin} = \frac{\lambda _{L}^{2}(0)\xi (0)}{2s}\mu _{0}I_{c}^{2}(0)(1 - \frac{T}{T_{c}})^{3/2}  \approx k_{B}T$$
at $T \approx  0.993 \ T_{c}$; this corresponds to the temperature at which the CVC hysteresis of the 18-ring system (Fig. 2) disappears.

\subsection{Magnetic-Field Dependence of Critical Currents $I_{c+}$ and $I_{c-}$}
Our measurements showed that the CVC portion with hysteresis, corresponding to the current decrease, cannot be the cause of the CVC asymmetry and periodic variation with the magnetic field. This portion is characterized by the absence of systematic asymmetry with respect to zero and appreciable variation with the magnetic field. In particular, the current $I_{c, s}$ at which the return to the superconducting state occurs is independent of both the measuring current direction and magnetic field. In contrast to this, the critical current corresponding to the transition to the resistive state depends on both the measuring current direction and magnetic field value. We denote these quantities measured at opposite $I_{ext}$ directions as $I_{c+}$ and $I_{c-}$ (Figs. 2 and 3). For the third type of CVC, $I_{c+}$ and $I_{c-}$ correspond to a sharp transition of the whole system to the resistive state (Fig. 2); for the second type of CVC, $I_{c+}$ and $I_{c-}$ correspond to the transition onset (Fig. 3). The technique for constructing the dependences of these quantities on the magnetic field $I_{c+}(B)$ and $I_{c-}(B)$ is described in Section 2.

Measurements performed at various temperatures on both single rings and systems of rings showed periodic dependences of the critical current $I_{c+}(B)$ and $I_{c-}(B)$ on the magnetic field. In all cases, the period $B_{0} = 1.4 \div 1.6 \ G$ corresponded to the flux quantum $B_{0}S = \Phi _{0} \approx  20.7 \ G \mu m^{2}$ inside a ring with an area of $S = 14.8 \div  13.0 \ \mu m^{2}$. Since simultaneous changes in the magnetic field $B$ and external current $I_{ext}$ directions are equivalent to a system rotation by 180°, which should not cause changes in the measured values, the equality $I_{c+}(B) = I_{c }(-B)$ should be satisfied for all values of $B$, where $B = B_{sol} + B_{res}$ ($B_{sol} = k_{sol}I_{sol}$) is the magnetic field induced by the copper solenoid, which was measured by the current $I_{sol}$ and the residual magnetic field $B_{res}$. For the measured dependences, the equality $I_{c+}(B) = I_{c }(-B)$ was satisfied at $B_{res} \approx  0.15 \ G$, which corresponded to the residual field obtained from other measurements. All figures show the dependences of the critical current and other quantities on $\Phi /\Phi _{0} = S(B_{sol} + B_{res})/\Phi _{0}$. 

\begin{figure}
\includegraphics{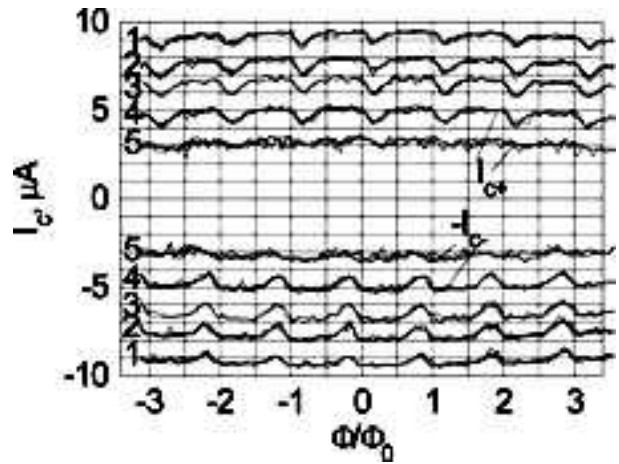}
\caption{\label{fig:epsart} Dependencies of critical currents on the magnetic field, $I_{c+}(\Phi /\Phi _{0})$ and $I_{c-}(\Phi /\Phi _{0})$, measured on the system of 18 asymmetric rings at temperatures (1) $T \approx  1.234 \ K \approx  0.972 T_{c}$, (2) $T \approx  1.238 \ K \approx  0.975T_{c}$, (3) $T \approx  1.241 \ K \approx  0.977T_{c}$, (4) $T \approx  1.245 \ K \approx  0.981T_{c}$, and (5) $T = 1.250 \ K \approx  0.985T_{c}$.}
\end{figure}

\begin{figure}
\includegraphics{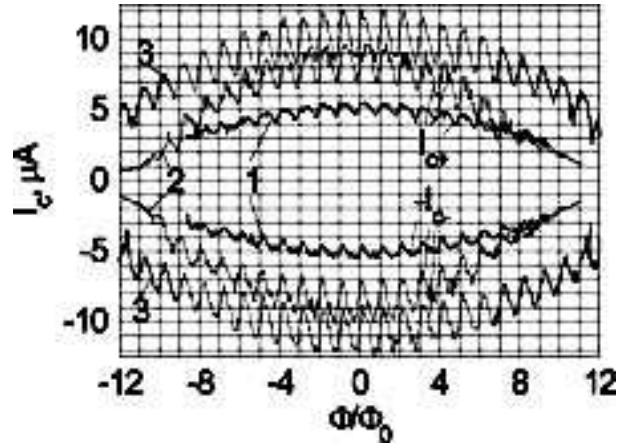}
\caption{\label{fig:epsart} Critical current oscillations $I_{c+}(\Phi /\Phi _{0})$ and $I_{c-}(\Phi /\Phi _{0})$ in a magnetic field, measured (1) on the system of 18 rings with $w_{con} = 0.4 \ \mu m$ at $T \approx  1.245 \ K \approx  0.981T_{c}$; (2) on the 20- ring system with $w_{con} = 1 \ \mu m$ at $T \approx  1.209 \ K \approx  0.984T_{c}$; and (3) on the single ring with $w_{con} = 0.6 \ \mu m$ at $T \approx  1.222 \ K \approx  0.991T_{c}$.}
\end{figure}

The noise level allowed us to reliably observe the periodic dependences $I_{c+}(\Phi /\Phi _{0})$ and $I_{c-}(\Phi /\Phi _{0})$ at a critical current level exceeding $3 \ \mu A$ (Fig. 5). Near $T_{c}$, where the first type of CVCs are observed, the dependencies $I_{c+}(\Phi /\Phi _{0})$ and $I_{c-}(\Phi /\Phi _{0})$ were not studied because of the small critical current $I_{c} < 3 \ \mu A$. At the same time, the constraint on the minimum temperature of about 1.19 K (which could be achieve on the setup used) did not allow us to observe oscillations at a critical current larger than $30 \ \mu A$. In addition to the periodic dependence, a monotonic decrease in the critical current with the magnetic field is observed (Fig. 6). Despite this, more than 25 oscillation periods were observed (Fig. 6). Measurements performed on systems of 20 rings connected in series with a current contact width $w_{con} = 0.4 \ \mu m$ and $w_{con} = 1 \ \mu m$ and on single rings with $w_{con} = 0.6 \ \mu m$ and $w_{con} = 0.7 \ \mu m$ showed that there is no qualitative difference between the dependencies $I_{c+}(\Phi /\Phi _{0})$ and $I_{c-}(\Phi /\Phi _{0})$ obtained on the ring system and the single ring (Fig. 6). At the same time, a certain effect of the current contact width $w_{con}$ is observed. The critical current decreases with increasing magnetic field to a greater extent in a structure with wider current contacts (Fig. 6). We recall here that other parameters, i.e., the width of ring halves $w_{w} = 0.4 \ \mu m$, $w_{n} = 0.2 \ \mu m$, and the ring diameter $2r = 4 \ \mu m$, were identical in all structures. We should emphasize the qualitative difference between the dependences $I_{c+}(\Phi /\Phi _{0})$ and $I_{c-}(\Phi /\Phi _{0})$ measured on structures with current contact widths smaller and larger than (or equal to) the total width of ring halves, $w_{con} = 0.4 \ \mu m < w_{w} + w_{n} = 0.6 \ \mu m$  and $w_{con} = 1, 0.7, 0.6 \ \mu m \geq w_{w} + w_{n} = 0.6 \ \mu m$ (Fig. 6). In the former case, there are minima and plateaus in the dependencies $I_{c+}(\Phi /\Phi _{0})$ and $I_{c-}(\Phi /\Phi _{0})$ (Figs. 5 and 6); in the latter case, there are minima and maxima (Figs. 6 and 9).

The critical current anisotropy 
$$I_{c,an}(\Phi /\Phi _{0}) = I_{c+}(\Phi /\Phi _{0}) - I_{c-}(\Phi /\Phi _{0})$$
is a periodic sign-alternating function of the magnetic field (Fig.7). In all cases studied, this dependence crosses the zero value at $\Phi  = n\Phi _{0}$ and $\Phi  = (n+0.5)\Phi _{0}$ (Fig. 7). We detected that the dependences of the critical current, measured in opposite directions, are similar or almost similar, i.e., coinciding, $I_{c-}(\Phi /\Phi _{0}) \approx  I_{c+}(\Phi /\Phi _{0} + \Delta \phi )$, when one of the dependencies is shifted along the magnetic field axis by  $\Delta \phi  =  \Delta \Phi /\Phi _{0} \leq 0.5$ (Fig. 8). The critical current anisotropy
$$I_{c,an}(\Phi /\Phi _{0}) = I_{c+}(\Phi /\Phi _{0}) - I_{c+}(\Phi /\Phi _{0} + \Delta \phi)$$
and all associated effects are due to this shift. The periodicity of the dependencies $I_{c+}(\Phi /\Phi _{0})$ and $I_{c-}(\Phi /\Phi _{0})$ and a large number of periods, which can be observed, allow sufficiently accurate determination of $\Delta \phi $ by superposing the measured dependencies $I_{c-}(\Phi /\Phi _{0})$ and $I_{c+}(\Phi /\Phi _{0} + \Delta \phi )$. Since these dependences were measured simultaneously, i.e., $I_{c+}$ and  $I_{c-}$  were measured sequentially for each value of $\Phi /\Phi _{0}$, the uncertainty in the $\Phi /\Phi _{0}$ magnitude cannot affect the determination accuracy of the relative shift  $\Delta \phi $. Therefore, a  $\Delta \phi \Phi _{0}$ determination accuracy of up to $0.02\Phi _{0}$ can be guaranteed.

The dependences $I_{c+}(\Phi /\Phi _{0})$ and $I_{c-}(\Phi /\Phi _{0})$ measured at various temperatures are similar (see Figs. 5 and 18), and the shift  ö is independent of temperature. The shift of the dependencies measured on single rings is equal to half the flux quantum,  $\Delta \phi  = 0.50 \pm  0.02$ for a ring with $w_{con} = 0.6 \ \mu m$ and  $\Delta \phi  = 0.48 \pm  0.05$ for a ring with $w_{con} = 0.7 \ \mu m$. A smaller shift is observed for the dependencies measured on 20-ring systems. We now cannot argue with certainty that there is a principal difference between specifically the single ring and ring system. On the system with $w_{con} = 1 \ \mu m \geq w_{w} + w_{n} = 0.6 \ \mu m$, for which the dependencies $I_{c+}(\Phi /\Phi _{0})$ and $I_{c-}(\Phi /\Phi _{0})$ qualitatively does not differ from the dependencies obtained for single rings, we performed only a single measurement which yielded  $\Delta \phi  = 0.30 \pm  0.02$. On the system with $w_{con} = 0.4 \ \mu m$, we carried out detailed measurements at various temperatures (Figs. 5 and 8), which showed  $\Delta \phi  = 0.36 \pm  0.02$. However, in this case, it is impossible with certainty to call this the cause in the difference in the shift from  $\Delta \phi  = 0.50 $; i.e., is it associated with the number of rings or with the qualitative difference of the dependencies $I_{c+}(\Phi /\Phi _{0})$ and $I_{c-}(\Phi /\Phi _{0})$ measured on structures with $w_{con} < w_{w} + w_{n} $ and $w_{con} \geq w_{w} + w_{n}$ (Fig. 6).

It is clear that the observed periodicity of the dependencies $I_{c+}(\Phi /\Phi _{0})$ and $I_{c-}(\Phi /\Phi _{0})$ of the critical current on the magnetic field is associated with quantization of the velocity of superconducting pairs in the ring (see (1)). As the ring returns to the superconducting state, the quantum number $n$ defining the momentum circulation of superconducting pairs takes an integer value. The quantity $n - \Phi /\Phi _{0}$ nonzero at $\Phi \neq n\Phi _{0}$ defines the relation between the velocities $v_{sn}$ and  $v_{sw}$ in ring halves according to (1). According to this relation, 
$$l_{n}v_{sn} - l_{w}v_{sw} = 0.5l(v_{sn} - v_{sw}) = \frac{2\pi \hbar }{m}(n - \frac{\Phi }{\Phi _{0}})$$
at the measuring current
$$I_{ext} = I_{n} + I_{w} = s_{n}j_{n} + s_{w}j_{w} = 2en_{s}(s_{n}v_{sn} + s_{w}v_{sw})$$
the velocities should be equal: in the narrow ring half and
$$v_{sn} = \frac{I_{ext}}{2en_{s}(s_{n} + s_{w})} + \frac{2\hbar }{mr}\frac{s_{w}}{s_{n}+s_{w}}(n - \frac{\Phi }{\Phi _{0}})$$
in the wide ring half,
$$v_{sw} = \frac{I_{ext}}{2en_{s}(s_{n} + s_{w})} - \frac{2\hbar }{mr}\frac{s_{n}}{s_{n}+s_{w}}(n - \frac{\Phi }{\Phi _{0}})$$
The measured critical currents given in Section 3.1 suggest that these values correspond to the depairing velocities $v_{sc}$, which would be expected for a narrow superconducting channel \cite{Tinkham}. Therefore, the critical current $I_{c+}(\Phi /\Phi _{0})$, $I_{c-}(\Phi /\Phi _{0})$ should correspond to the measuring current $|I_{ext} |$ at which the superconducting pair velocity reaches the depairing velocity in one of structural elements. If the velocity reaches a critical value in the narrow ring half, $|v_{sn} | = v_{sc}$ , 
$$I_{c+}, I_{c-} = I_{c0} - 2I_{p,A}|n - \frac{\Phi }{\Phi _{0}}|(1 + \frac{s_{w}}{s_{n}}) \eqno{(2a)}$$
if this occurs in the wide half, $|v_{sw} | = v_{sc}$ ,
$$I_{c+}, I_{c-} = I_{c0} - 2I_{p,A}|n - \frac{\Phi }{\Phi _{0}}|(1 + \frac{s_{n}}{s_{w}}) \eqno{(2b)}$$
where $I_{c0} = 2en_{s}(s_{n} + s_{w})v_{sc}$ is the critical current at $\Phi /\Phi _{0} = n$; $I_{p, A} = 2en_{s}(\hbar /mr)s_{n}s_{w}/(s_{n} + s_{w})$ is the value corresponding to the persistent current amplitude as $n -  \Phi /\Phi _{0}$ varies between -0.5 and 0.5. According to the geometry shown in Fig. 1, expression (2a) is applicable for $I_{c+}$ at positive $n -  \Phi /\Phi _{0}$ and for $I_{c-}$ at negative values; this is vice versa for expression (2b). Relations (2) were obtained on then basis of the conventional description of the superconducting state as a macroscopic quantum phenomenon in the approximation of small width and cross section of ring halves, $w_{n}, w_{w} \ll  2r$; $s_{n}, s_{w} < \lambda _{L}^{2}(T)$. We proceeded from velocity quantization expression (1) which follows from the Bohr quantization of the momentum circulation of superconducting pairs
$$\oint_{l}dlp = \oint_{l}dl (mv_{s} +  2eA) = m\oint_{l}dl v + 2e\Phi = n 2\pi \hbar $$
Expressions (2) can also be derived from the Ginzburg-Landau theory, starting from the expression for the superconducting current density $j_{s} = (2e/m)n_{s}(\hbar \bigtriangledown \varphi  - 2eA)$ \cite{Tinkham}, where $\varphi $ is the wavefunction phase and $A$ is the vector potential.

\begin{figure}
\includegraphics{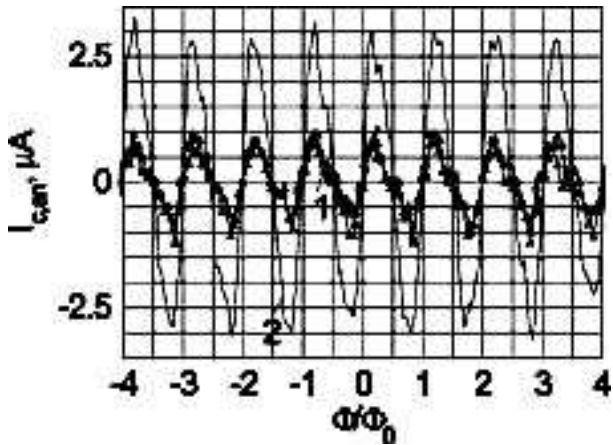}
\caption{\label{fig:epsart} Critical current anisotropy oscillations $I_{c,an}(\Phi /\Phi _{0})$ measured (1) on the system of 18 rings with $w_{con} = 0.4 \ \mu m$ at $T \approx  1.247 \ K \approx  0.982T_{c}$ and (2) on the single ring with $w_{con} = 0.6 \ \mu m$ at $T \approx  1.221 \ K \approx  0.991T_{c}$. }
\end{figure}

\begin{figure}
\includegraphics{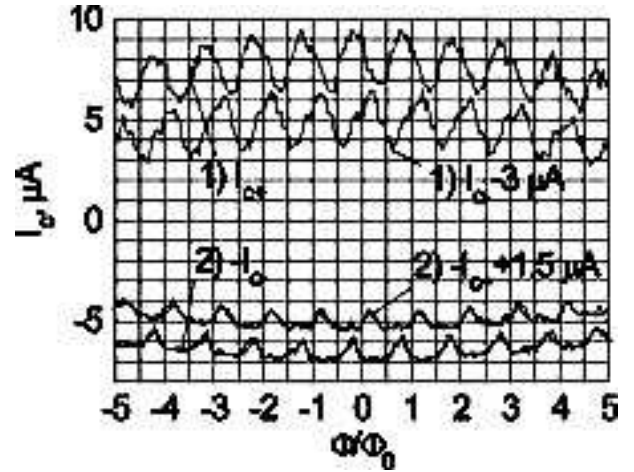}
\caption{\label{fig:epsart} Relative shift of oscillations $I_{c+}(\Phi /\Phi _{0})$ and $I_{c-}(\Phi /\Phi _{0})$ along the $\Phi /\Phi _{0}$ axis, measured at opposite directions of the measuring current (1) on the 20-ring system with $w_{con} = 1 \ \mu m$ at $T \approx  1.209 \ K \approx  0.984T_{c}$ and (2) on the 18-ring system with $w_{con} = 0.4 \ \mu m$ at $T \approx  1.241 \ K \approx  0.978T_{c}$.}
\end{figure}

Let us consider those aspects of the measured dependencies $I_{c+}(\Phi /\Phi _{0})$ and $I_{c-}(\Phi /\Phi _{0})$ obtained according to the generally accepted concepts on the superconducting state, which are consistent and inconsistent with relations (2). The first and main consistency is the periodicity of expected (2) and observed dependencies $I_{c+}(\Phi /\Phi _{0})$ and $I_{c-}(\Phi /\Phi _{0})$. Since the energy difference between allowed levels defined by the quantum number $n$ is large (much larger than the thermal fluctuation energy $k_{B}T$), the lowest level, i.e., the level with the smallest value of $(n - \Phi /\Phi _{0})^{2}$, has an overwhelming probability. Therefore, the equilibrium quantity $|n - \Phi /\Phi _{0}|$ in expressions (2) is a periodic function of the magnetic field  $\Phi  = SB$ with period $\Phi _{0}$. Therefore, the periodic dependencies $I_{c+}(\Phi /\Phi _{0})$ and $I_{c-}(\Phi /\Phi _{0})$ (Figs. 5 and 6) are observed.

The second consistency is the qualitative difference between the dependencies $I_{c+}(\Phi /\Phi _{0})$ and $I_{c-}(\Phi /\Phi _{0})$ measured on structures with $w_{con} < w_{w} + w_{n} $ and $w_{con} \geq w_{w} + w_{n}$ (Figs. 5 and 6). In the structure with cross section $s_{con} = w_{con}d \geq s_{n} + s_{w} = (w_{w} + w_{n})d$, when $I_{ext}/(s_{n} + s_{w}) \geq  I_{ext}/s_{con}$, the current density $j_{s,con} = I_{ext}/s_{con}$ in current contacts cannot exceed the current density simultaneously in both halves, since $I_{ext} = s_{n}j_{n} + s_{w}j_{w}$. In this case, the velocity reaches a critical value initially in one of them; then, one of expressions (2) is valid. In contrast to this case, in the structure with $s_{con} < s_{n} + s_{w}$, the critical velocity can be reached initially in current contacts at $|I_{ext} | = 2en_{s}s_{con}v_{sc} < I_{c0} = 2en_{s}(s_{n} + s_{w})v_{sc}$. This quantity is independent of $|n - \Phi /\Phi _{0} |$; therefore, the measured critical current should not depend on the magnetic field, when $|I_{ext} | = 2en_{s}s_{con}v_{sc}$ is smaller than the critical current (2) controlled by the velocity in one of ring halves. This corresponds to the existence of the plateau with $I_{c+} = I_{c-} = I_{c, pl} = 2en_{s}s_{con}v_{sc}$ in the dependencies $I_{c+}(\Phi /\Phi _{0})$ and $I_{c-}(\Phi /\Phi _{0})$ measured on the structure with $w_{con} = 0.4 \ \mu m < w_{w} + w_{n} = 0.6 \ \mu m$ (Figs. 5 and 6) and the absence of plateau for the structures with $w_{con} = 1; 0.7; 0.6 \ \mu m \geq w_{w} + w_{n} = 0.6 \ \mu m$  (Fig. 6).

A decrease in the critical current with increasing magnetic field (Fig. 6) is also consistent with commonly accepted concepts. This is associated with a decrease in the superconducting pair density $n_{s}$ in finite-width strips \cite{Tinkham}. The observed decrease not only in $I_{c+}(\Phi /\Phi _{0})$ and $I_{c-}(\Phi /\Phi _{0})$, but also in their oscillation amplitude at large $\Phi /\Phi _{0}$ (Fig. 6) is consistent with (2), since not only the critical $I_{c0}$ current, but also the persistent $I_{p, A}$ current depend on the pair density $n_{s}$. The effect of the current contact width $w_{con}$ on decreases in $I_{c0}$ at large $\Phi /\Phi _{0}$ in structures with $w_{con} \geq w_{w} + w_{n}$ (Fig. 6) is a manifestation of the proximity effect: stronger suppression of $n_{s}$ in wider current contacts results in $n_{s}$ suppression in rings through the proximity effect. The results of our measurements suggest that the proximity effect appears at distances of several micrometers. In a structure with contact areas wider than $4 \ \mu m$ arranged at a distance of $2 \ \mu m$ from the ring, the superconductivity is suppressed even in weak fields and no more than five oscillation periods can be observed. More than 25 periods of critical current oscillations can be observed on such a structure, but with areas spaced from the ring at a distance longer than $10 \ \mu m$ (Fig. 6).

The fourth consistency is the periodic sign-alternating dependence of the critical current anisotropy on the magnetic field. According to (2), for structures with $w_{con} \geq w_{w} + w_{n}$, the anisotropy is given by
$$I_{c,an} =  I_{c+} - I_{c-} = 2I_{p,A}(n - \frac{\Phi }{\Phi _{0}})(\frac{s_{w}}{s_{n}} - \frac{s_{n}}{s_{w}}) \eqno{(2c)}$$
The sign of the anisotropy is positive at positive $n - \Phi /\Phi _{0}$, according to the geometry shown in Fig. 1. As positive, we take the direction from right to left for $I_{ext}$ and clockwise for the persistent current,
$$I_{p} = I_{p,A}2(n - \Phi /\Phi _{0})$$
The amplitude $I_{A,an}$ of observed anisotropy oscillations $I_{c,an}(T,\Phi /\Phi _{0}) = I_{A,an}(T)f_{an}(\Phi /\Phi _{0})$ linearly increases as it recedes from $T_{c}$, $ I_{A,an}(T) \propto 1 -T/T_{c}$, according to
$$I_{p,A} = n_{s} 2e\frac{s_{n}s_{w}}{s_{n}+s_{w}}\frac{\hbar }{mr} \propto n_{s} \propto (1 - \frac{T}{T_{c}})$$
The value of $I_{A,an}$ is close to the expected value of $(s_{w}/s_{n} - s_{n}/s_{w})I_{p, A}$ (see Section 3.5 for more detail). 

Having regard for such important consistencies, the measured dependencies $I_{c+}(\Phi /\Phi _{0})$ and $I_{c-}(\Phi /\Phi _{0})$ differ essentially from the expected dependences (2). On account of the importance of these inconsistencies, we consider them in separate Section 3.6. A certain paradox is that, despite these fundamental differences, both expected (2c) and observed (Fig.7), critical current anisotropy oscillations $I_{c,an}(\Phi /\Phi _{0})$ explain the rectified voltage oscillations $V_{dc}(\Phi /\Phi _{0})$ observed in \cite{QuOs2003}. In the Section 3.3, we shall consider the consistency between measured oscillations$I_{c+}(\Phi /\Phi _{0})$, $I_{c-}(\Phi /\Phi _{0})$ and $V_{dc}(\Phi /\Phi _{0})$. In Section 3.4, we present the measured temperature dependencies of quantities characterizing the rectification efficiency. It should be emphasized that these dependencies are consistent with the observed critical current oscillations with temperature. To more emphasize the strangeness of the inconsistency between expected and observed oscillations $I_{c+}(\Phi /\Phi _{0})$, $I_{c-}(\Phi /\Phi _{0})$, we show in Section 3.5 that the observed values and temperature dependencies of the amplitude $I_{c,an}(\Phi /\Phi _{0})$ are consistent with expected dependencies of persistent current.

\subsection{Quantum Oscillations of ac Voltage as a Result of Alternating Current Rectification}
The observed asymmetry of the critical current and its periodic dependence on the magnetic flux, $I_{c,an}(\Phi /\Phi _{0}) = I_{c+}(\Phi /\Phi _{0}) - I_{c-}(\Phi /\Phi _{0})$, allow us to explain the rectification effect and the dependence $V_{dc}(\Phi /\Phi _{0})$ of the rectified voltage on the magnetic flux. A comparison of the dependencies $I_{c+}(\Phi /\Phi _{0})$ and $I_{c-}(\Phi /\Phi _{0})$ with dependences $V_{dc}(\Phi /\Phi _{0})$ caused by both the sinusoidal, $I_{ext}(t) = I_{0}sin(2\pi ft)$, and sawtooth currents showed that $|V_{dc}|(\Phi /\Phi _{0})$ maxima are close to $I_{c+}(\Phi /\Phi _{0})$ and $I_{c-}(\Phi /\Phi _{0})$ minima in all cases (Fig. 9). The position of $|V_{dc}|(\Phi /\Phi _{0})$ maxima shifts from the values closer to integer flux quanta, $\Phi  = (n + 0.14)\Phi _{0}$ and $\Phi  = (n + 0.86)\Phi _{0}$, at small alternating current amplitudes to the values $\Phi  = (n + 0.25)\Phi _{0}$ and $\Phi  = (n + 0.75)\Phi _{0}$ corresponding to the midpoint between the flux quantum and half the quantum at large amplitudes $I_{ext}$. Good similarity is observed between quantum oscillations of the dc voltage and quantum oscillations of the critical current anisotropy $V_{dc}(\Phi /\Phi _{0}) \propto -I_{c,an}(\Phi /\Phi _{0})$ (Fig. 10).

The rectified voltage, i.e., the time-averaged voltage
$$V_{dc} = \Theta ^{-1}\int_{\Theta } dtV(I_{ext}(t))$$
can be calculated from the $V(I_{ext})$ CVC (see Figs. 2-4) and the time dependence of the external current, e.g., $I_{ext}(t) = I_{0}sin(2\pi ft)$. The measurement results of the CVC and the magnetic dependences of the critical current anisotropy, presented in Sections 3.1 and 3.5, respectively, allow quantitative description of the rectification effect resulting in observation of quantum oscillations $V_{dc}(\Phi /\Phi _{0})$. The rectified voltage oscillation amplitude depends on several factors, including the CVC type, but the main one is the relation between the external current amplitude $I_{0}$ and the critical currents $I_{c+}(\Phi /\Phi _{0})$ and $I_{c-}(\Phi /\Phi _{0})$. 

\begin{figure}
\includegraphics{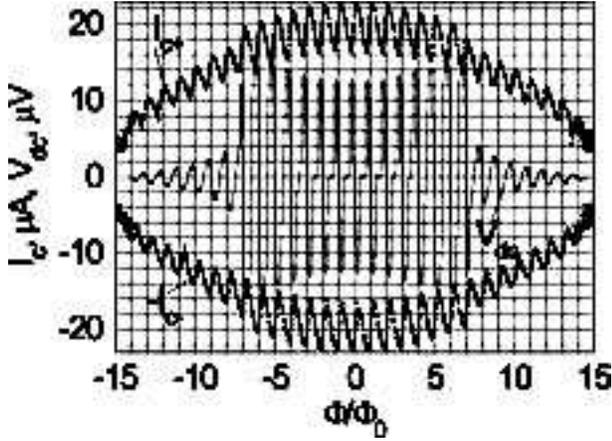}
\caption{\label{fig:epsart} Comparison of oscillations of the critical current $I_{c+}(\Phi /\Phi _{0})$, $I_{c-}(\Phi /\Phi _{0})$ and the dc voltage $V_{dc}(\Phi /\Phi _{0})$ induced by alternating current with frequency $f = 400 \ Hz$ and an amplitude of $I_{0} = 17.6 \ \mu A$ on the single ring at $T \approx  1.216 \ K \approx  0.986T_{c}$. }
\end{figure}

\begin{figure}
\includegraphics{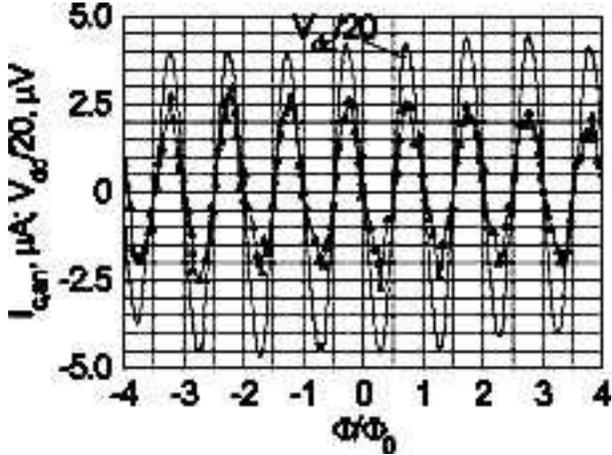}
\caption{\label{fig:epsart} Comparison of oscillations of the critical current anisotropy $I_{c,an}(\Phi /\Phi _{0})$ and the dc voltage $V_{dc}(\Phi /\Phi _{0})$ induced by alternating current with frequency $f = 500 \ Hz$ and an amplitude of $I_{0} = 10.6 \ \mu A$ in a magnetic field on the 20- ring system at $T \approx  1.209 \ K \approx  0.985T_{c}$.}
\end{figure}

The dc voltage $V_{dc}(\Phi /\Phi _{0})$ appears when the external current amplitude $I_{0}$ exceeds the least $I_{c,min}$ among critical values of $I_{c+}(\Phi /\Phi _{0})$ and $I_{c-}(\Phi /\Phi _{0})$. At larger $\Phi /\Phi _{0}$, this occurs at smaller amplitudes $I_{0}$ (Fig. 11), since the critical current is here suppressed by the magnetic field (see Figs. 6 and 9). A decrease in the amplitude $V_{A}$ with increasing $I_{0}$ immediately after reaching the maximum $V_{A, max}$ at $I_{0} = I_{0, max}$ (Figs. 11 and 12) is caused by the opposite-sign voltage appeared as $I_{0}$ reaches the larger value $I_{c, max}$ among $I_{c+}(\Phi /\Phi _{0})$ and $I_{c-}(\Phi /\Phi _{0})$. The peak width in the dependence $V_{A}(I_{0})$ (Fig. 12) corresponds to the $I_{c+}(\Phi /\Phi _{0})$, $I_{c-}(\Phi /\Phi _{0})$ amplitude (Figs. 5 and 6) at the third-type CVC. For this CVC type with stepwise variation and strong hysteresis (Figs. 2 and 4), a stepwise increase in the rectified voltage would be expected at $I_{0} =I_{c+}(\Phi /\Phi _{0}), I_{c+}(\Phi /\Phi _{0})$. The absence of such a step is probably associated with uncontrollable noises $I_{noise}$, due to which the actual external current $I_{ext}(t) = I_{0}sin(2\pi ft) + I_{noise}(t)$ and its actual amplitude vary in time and not always coincides with $I_{0}$.

\begin{figure}
\includegraphics{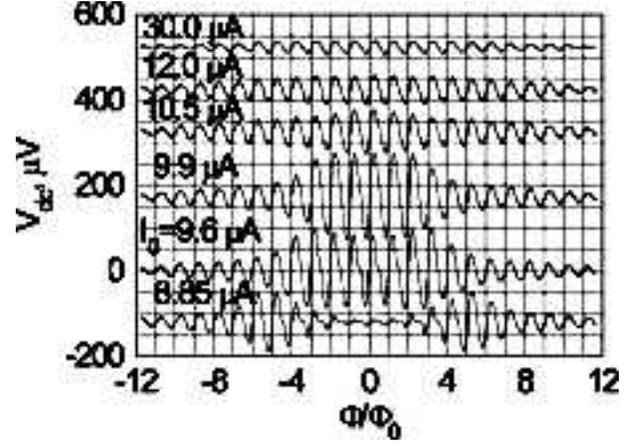}
\caption{\label{fig:epsart} Oscillations of dc voltage $V_{dc}(\Phi /\Phi _{0})$ induced by alternating current with frequency $f = 40 \ kHz$ and various amplitudes $I_{0} = 8.85, 9.6, 9.9, 10.5, 12$, and $30 \ \mu A$ on the 18-ring system at $T \approx  1.234 \ K \approx  0.972T_{c}$. All the dependences, except for $I_{0} = 9.6 \ \mu A$, are shifted vertically. }
\end{figure}

\begin{figure}
\includegraphics{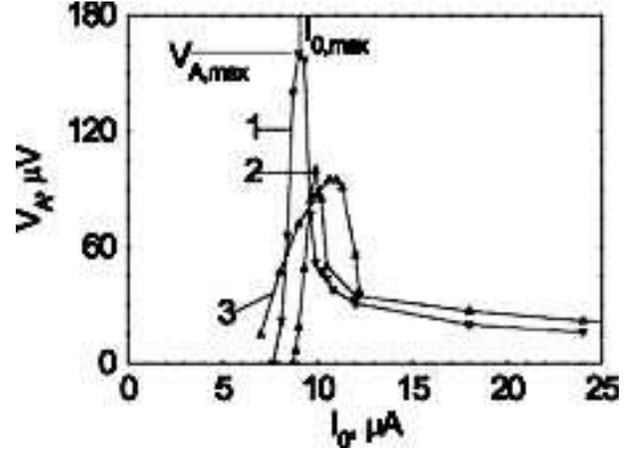}
\caption{\label{fig:epsart} Dependences of the amplitude $V_{A}$ of rectified voltage oscillations $V_{dc}(\Phi /\Phi _{0})$ on the external current amplitude $I_{0}$ for the 18-ring system with $w_{con} = 0.4 \ \mu m$ at (1) sinusoidal and (2) sawtooth currents with $f = 40 \ kHz$, $T \approx  1.234 \ K \approx  0.972T_{c}$ and (3) the 20-ring system with $w_{con} = 1 \ \mu m$ at sinusoidal current with $f = 1 \ kHz$, $T \approx  1.214 \ K \approx  0.989T_{c}$.}
\end{figure}

\begin{figure}
\includegraphics{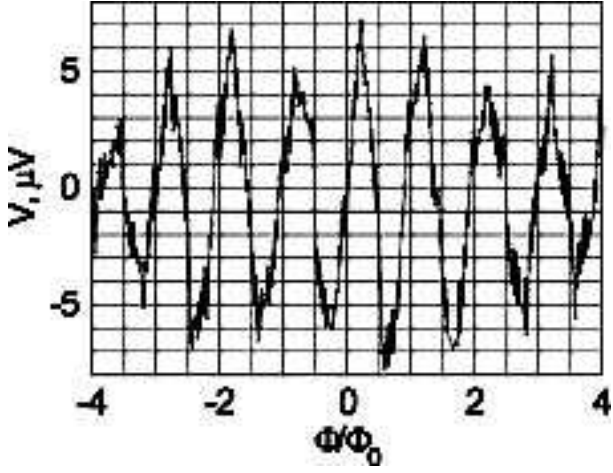}
\caption{\label{fig:epsart} Quantum oscillations of the dc voltage induced by sawtooth alternating current with frequency $f = 40 \ kHz$ and an amplitude of $I_{0} = 3.2 \ \mu A$ on the 18-ring system at temperature $T \approx  1.263 \ K \approx  0.996T_{c}$, corresponding to the first-type CVC, see Fig. 2. }
\end{figure}

The dependence $V_{A}(I_{0})$ shape is unchanged with the external current $I_{ext}(t)$ frequency and shape (sinusoidal or sawtooth) and the CVC type (Fig. 12). However, the $V_{A}(I_{0})$ peak height and width depend on both the current $I_{ext}(t)$ shape and CVC type (Fig. 12). As expected, the maximum amplitude $V_{A, max}$ is observed at the third type CVC and sinusoidal current (Fig. 12). For the second- type CVC, the $V_{A}(I_{0})$ peak is wider than that for the third-type CVC (Fig. 12).

In the presence of the CVC hysteresis, the voltage $V(I_{ext}(t))$ has a nonzero value for a relatively large fraction of the $I_{ext}(t)$ period, even if the current amplitude $I_{0}$ only insignificantly exceeds the critical current $I_{c+}(\Phi /\Phi _{0})$, $I_{c-}(\Phi /\Phi _{0})$. Therefore, the hysteresis increases the rectification efficiency. In the approximation $V \approx  R_{n}I_{ext}(t)$, as the sawtooth current $I_{ext}(t)$ decreases down to the superconducting transition at $I_{ext} = I_{c,s}$ (Figs. 2 and 4), the maximum amplitude of the rectified voltage, observed at $I_{0} \geq  I_{0, max}$, can be estimated by the formula
$V_{A,max} \approx 0.25R_{n}(2I_{c,max} - I_{c,min} - I_{c,s}) \approx 0.25R_{n}(I_{c,max} - I_{c,s} + I_{A,an})$.
For all the measured structures with the third-type CVC, the hysteresis value is larger than the anisotropy value, $I_{c,max} - I_{c,s} > I_{A,an}$, even taking into account that $I_{c,s}$ should be considered as an effective value corresponding to the hysteresis area (Fig. 2). The anisotropy amplitude of the critical current, measured, e.g., at $T \approx 1.234 \ K \approx  0.972T_{c}$ on the 18-ring system with normal state resistance $R_{n} \approx  92 \ \Omega $, is $I_{A,an} \approx  1 \ \mu A$. This corresponds to $V_{A,max} \approx 0.25 R_{n}I_{a,an} \approx 23 \ \mu V$ in the absence of hysteresis, which is several times smaller than the maximum amplitude $V_{A,max} \approx  100 \ \mu V$ of quantum oscillations which induce a sawtooth current at this temperature. Estimations taking into account the hysteresis $I_{c,max} - I_{c,s}$ yield $V_{A,max}$ values close to the measured values. 

In addition to the quantitative difference, there is also an important qualitative difference between quantum oscillations $V_{dc}(\Phi /\Phi _{0})$ observed in the presence (Fig. 11) and in the absence of CVC hysteresis (Fig. 13). The existence of the hysteresis means that not only structure regions in which the current density exceeds the critical value ($j > j_{c}$) can be in the resistive state, but also neighboring regions in which $j < j_{c}$. In this case, the persistent current promotes the transition to the resistive state of not only that ring half in which the persistent current increases the current density, e.g., $j_{n} = I_{ext}/(s_{n} + s_{w}) + I_{p}/s_{n} > j_{c}$, but also in that ring half in which it decreases the current density, $j_{w} = I_{ext}/(s_{n} + s_{w}) - I_{p}/s_{w} < j_{c}$. The absence of CVC hysteresis means that the transition to the resistive state of one ring half does not result in a similar transition of the second half until the current density in it reaches the critical value.

\subsection{Temperature Dependence of the Maximum Amplitude $V_{A,max}$ of Quantum Oscillations. Rectification Efficiency}
It is known that the critical $I_{c}$ and persistent $I_{p}$ currents decrease when approaching the critical temperature $T_{c}$. It can be expected that the critical amplitude $I_{0c}\approx  I_{c}(T)$ of the current and the maximum amplitude $V_{A,max}$ of quantum oscillations $V_{dc}(\Phi /\Phi _{0}) = R_{as}I_{p}(\Phi /\Phi _{0})$ should decrease when approaching $T_{c}$ similarly to $I_{c}(T)$ and $I_{p}(T)$. A decrease in the critical amplitude $I_{0c}$ to zero near $T_{c}$ means that noises with arbitrarily low intensity can induce quantum oscillations $V_{dc}(\Phi /\Phi _{0})$ in this temperature range.

Here we present the first results of measurements of the temperature dependencies of the critical amplitude $I_{0c}(T)$, maximum amplitude $V_{A,max}(T)$ of quantum oscillations, and the external current amplitude $I_{0, max}(T)$ at which $V_{A,max}(T)$ is observed (Fig. 12). The results obtained show the difference of rectification mechanisms in the immediate vicinity of $T_{c}$ and at lower temperatures and make it possible to more reliably estimate the possibility to observe noises of arbitrarily low intensity.

\begin{figure}
\includegraphics{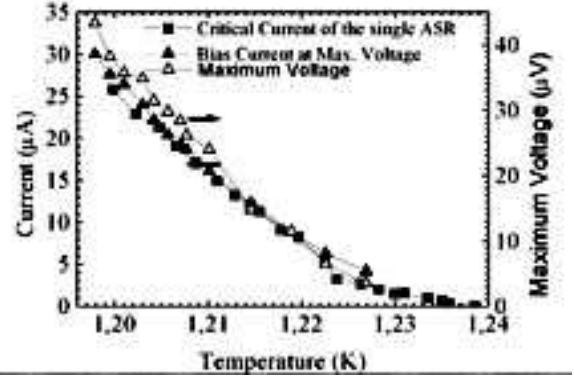}
\caption{\label{fig:epsart} Temperature dependences of the critical current $I_{c}$ (squares), the maximum amplitude $V_{A,max}$ of quantum oscillations of dc voltage, and the amplitude $I_{0,max}$ (closed triangles) of the external current inducing $V_{A,max}$ (open triangles), measured on the single ring.}
\end{figure}

\begin{figure}
\includegraphics{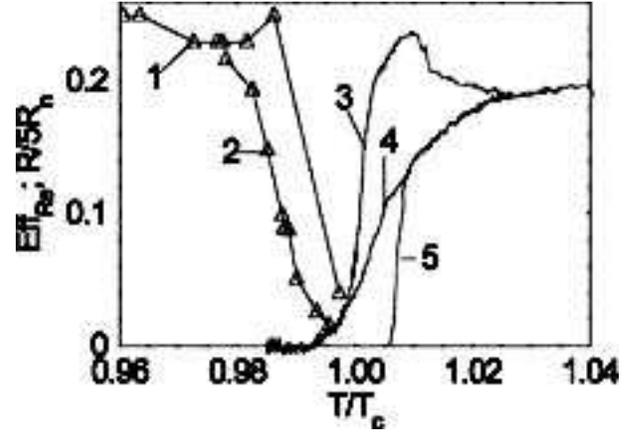}
\caption{\label{fig:epsart} Temperature dependences of the rectification efficiency $Eff_{Re} = V_{A,max}/I_{0,max}R_{n}$, measured on (1) the single ring and (2) the 18-ring system with $R_{n} = 92 \ \Omega $. The resistive transitions of (3) the single ring with $R_{n} = 3.3 \ \Omega $, (4) the 20- ring system with $R_{n} = 58 \ \Omega $, and (5) the 2-ring system with $R_{n} = 5.0 \ \Omega $. }
\end{figure}

The measurements performed on single rings showed that, as expected, $I_{0c}(T)$ and $I_{0,max}(T)$ are close to the dependence of the critical current $I_{c}(T)$ (Fig. 14). In this figure, only $I_{0,max}(T)$ is shown, since $I_{0c}(T)$ is analogous to it. The maximum amplitude $V_{A,max}(T)$ of quantum oscillations decreases with increasing $T$ similarly to $I_{0,max}(T)$ at temperatures not close to the resistive transition. In the range $T = 0.95 \div 0.99T_{c}$, the ratio $V_{A,max}/I_{0,max} = 0.79 \pm  0.03 \ \Omega $  (at the normal-state ring resistance $R_{n} \approx  3.3 \ \Omega $ ) is independent of temperature to within the measurement error and sharply decreases before the resistive transition (Fig. 15). Our measurements confirmed the increase in the amplitude of quantum oscillations with the number of rings, observed in \cite{QuOs2003}. 

To compare the rectification efficiency in systems with different numbers of rings $N$, the ratio $R_{eff} = V_{A,max}/I_{0,max}N$ can be used, as well as $Eff_{Re} = V_{A,max}/I_{0,max}R_{n}$, since the normal-state resistance $R_{n}$ increases with the number of rings. Our measurements showed that $R_{eff}$ and $Eff_{Re}$ do not significantly differ for a single ring and a system of rings at low temperatures. For the single ring, $R_{eff} \approx  0.79 \ \Omega $  and $Eff_{Re} \approx  0.24$ at $T = (1.18 \div 1.23) \ K = (0.95 \div 0.99)T_{c}$; for the system of 18 rings, $R_{eff} \approx  1.1 \ \Omega $  and $Eff_{Re} \approx  0.22$  at $T \approx  1.240 \ K \approx  0.977T_{c}$ (Fig. 15); for the 20-ring system, $R_{eff} \approx  0.45 \ \Omega $  and $Eff_{Re} \approx  0.16$  at $T \approx  1.210 \ K \approx  0.978T_{c}$. The high rectification efficiency $Eff_{Re}$ is observed at low temperatures due to the CVC hysteresis. For the second- and third-type CVCs with hysteresis, a higher efficiency is observed in the latter case. This is particularly clear when comparing the values of $Eff_{Re}$ measured on the system of 18 rings ($w_{con} = 0.4 \ \mu m$) with the third-type CVC at low temperatures (Fig. 2) and the 20-ring system ($w_{con} = 1 \ \mu m$) with the second-type CVC (Fig. 3). The sharp decrease in the rectification efficiency $Eff_{Re}$ with increasing temperature (Fig. 15) coincides with the change in the CVC type (Figs. 2-4). Despite the decrease in the rectification efficiency near $T_{c}$, the observed increase in $NR_{eff} = V_{A,max}/I_{0,max}$ with the number of rings $N$ confirms the possibility of using systems with a large number of asymmetric superconducting rings as highly sensitive noise sensors \cite{QuanDet}.

\subsection{Amplitudes of Critical and Persistent Current Anisotropy Oscillation}
According to expressions (2) for the critical current, obtained in Section 3.2 on the basis of conventional concepts, the critical current anisotropy of a structure with $s_{w}/s_{n} = 2$ and $s_{con} \geq  s_{n} + s_{w}$ should be given by
$$I_{c,an} = (\frac{s_{w}}{s_{n}} - \frac{s_{n}}{s_{w}})I_{p} = 1.5I_{p,A}(T)2(n - \frac{\Phi }{\Phi _{0}})$$
According to this relation, anisotropy oscillations $I_{c,an}(\Phi /\Phi _{0})$ measured at various temperatures should be similar, $I_{c,an}(\Phi /\Phi _{0}) = I_{A,an}(T)f_{an}(\Phi /\Phi _{0})$, and their amplitude should linearly decrease when approaching the critical temperature, $I_{A,an} = I_{A,an}(0)(1 - T/T_{c})$. Our measurements confirmed similarities of anisotropy oscillations of the critical current measured at various temperatures (Fig. 16). This allows us to compare the temperature dependencies and amplitudes of observed and expected oscillations despite their qualitative difference.

\begin{figure}
\includegraphics{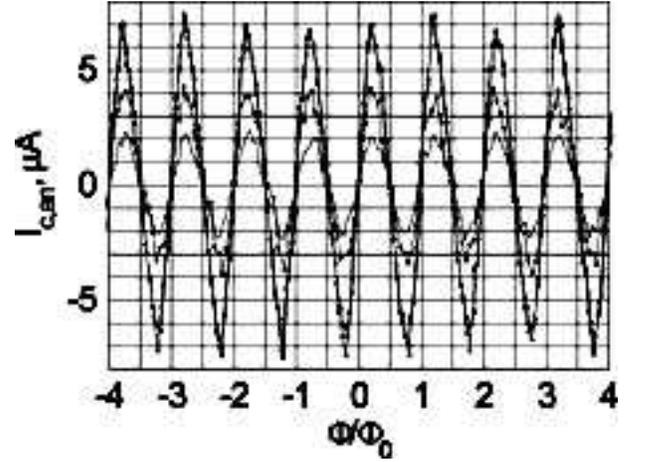}
\caption{\label{fig:epsart} Critical current anisotropy oscillations $I_{c,an}(\Phi /\Phi _{0})$, measured on the single ring with $w_{con} = 0.7 \ \mu m$ at various temperatures $T \approx  1.211 \ K \approx  0.972T_{c}$, $T \approx  1.227 \ K  \approx  0.983T_{c}$, and $T \approx  1.235 \ K \approx  0.992T_{c}$.}
\end{figure}

\begin{figure}
\includegraphics{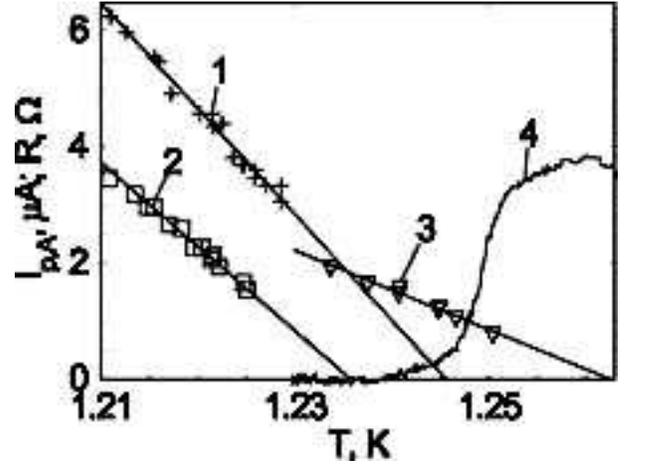}
\caption{\label{fig:epsart} Temperature dependences of $I_{A,an}/1.5$ and $I_{c,pl}/2 - I_{c+}/3$, which should correspond to the persistent current amplitude oscillations, measured on (1) the single ring with $w_{con} = 0.7 \ \mu m$, (2) the single ring with $w_{con} = 0.6 \ \mu m$, and (3) the 18-ring system with $w_{con} = 0.4 \ \mu m$. Symbols are experimental data. Lines are the dependences $I_{p,A}(T) = I_{p,A}(0)(1 - T/T_{c})$ for (1) $I_{p,A}(0) = 0.227 \ mA$ and $T_{c} = 1.246 \ K$, (2) $I_{p,A}(0) = 0.179 \ mA$ and $T_{c} = 1.236 \ K$, (3) $I_{p,A}(0) = 0.068 \ mA$ and $T_{c} = 1.263 \ K$, and (4) the resistive transition of the ring with $w_{con} = 0.7 \ \mu m$. }
\end{figure}

The expected amplitude of the critical current anisotropy of a structure with $s_{con} \geq  s_{n} + s_{w}$ should be given by $I_{A,an}(T) = 1.5I_{p,A}(T)$. Based on this relation, we can estimate the persistent current amplitude from measured amplitudes of the critical current anisotropy. For structures with $s_{con} <  s_{n} + s_{w}$, there is no simple relation between $I_{c,an}(T,\Phi /\Phi _{0})$ and $I_{p}(T,\Phi /\Phi _{0})$, since the critical current at an insufficiently large persistent current is controlled by its value in current contacts, $I_{c+} = I_{c-} = I_{c,pl} = s_{con} j_{c}$. The persistent current can be determined by the value by which it decreases the critical current $I_{c+}$ (or $I_{c-}$) (Fig. 5),
$$|I_{p}| = \frac{I_{c,pl}s_{n}}{s_{con}} - \frac{I_{c+}s_{n}}{s_{n}+s_{w}} = \frac{I_{c,pl}}{2} - \frac{I_{c+}}{3} $$
The persistent current amplitude calculated by these relations from experimentally determined $I_{A,an}(T)$ and $I_{c,pl}/2 - I_{c+}/3$ linearly decreases when approaching the critical temperature, $I_{p,A} = I_{p,A}(0)(1 - T/T_{c})$ (Fig. 17). This is consistent with the linear decrease in the persistent current amplitude
$$I_{p,A}(T) = 2en_{s}\frac{s_{n}s_{w}}{s_{n}+s_{w}}\frac{\hbar }{mr} = 2en_{s}\frac{2(s_{n}+s_{w})}{9}\frac{\hbar }{mr}   $$
proportional to the pair density $n_{s} \propto (1 - T/T_{c})$. Using the expression for the critical current density $j_{c} = n_{s} 2e(\hbar /m \surd 3\xi (T))$ \cite{Tinkham}, the persistent current amplitude can be expressed in terms of the critical current as
$$I_{p,A} = I_{c}(T)\frac{2}{9}\frac{\surd 3\xi (T)}{r} = I_{c}(0)\frac{2}{9}\frac{\surd 3\xi (0)}{r}(1 - \frac{T}{T_{c}})   $$
for $s_{con} \geq  s_{n} + s_{w}$ and
$$I_{p,A} =  I_{c}(0)\frac{1}{3}\frac{\surd 3\xi (0)}{r}(1 - \frac{T}{T_{c}})   $$
for $(s_{n} + s_{w})/s_{con} = 1.5$.

This allows us to compare the theoretical persistent currents and their values obtained from measurements of critical current oscillations, using the experimental temperature dependencies of the critical current given in Section 3.1,
$I_{c}(T) = I_{c}(0)(1 -T/T_{c})^{3/2}$,
with $I_{c}(0) \approx  5.5 \ mA$ and $4.3 \ mA$ for single rings with $w_{con} = 0.7 \ \mu m$ and $ 0.6 \ \mu m$, respectively, and $I_{c}(0) \approx  1.4 \ mA$ for the system of 18 rings with $w_{con} = 0.4 \ \mu m$.

The values $I_{p,A}(0) = 0.23 \ mA, 0.1 \ mA$, and $0.07 \ mA$ obtained from $I_{c+}(\Phi /\Phi _{0})$ and $I_{c-}(\Phi /\Phi _{0})$ measurements on rings with $w_{con} = 0.7 \ \mu m$ and $0.6 \ \mu m$ and the system of 18 rings with $w_{con} = 0.4 \ \mu m$ (Fig. 17) are close to the corresponding values $I_{c}(0)(2/9)(\surd 3 \xi (0)/r) = 2.1 \ mA(\xi (0)/r)$, $I_{c}(0)(2/9)(\surd 3 \xi (0)/r) = 1.6 \ mA(\xi (0)/r)$, and $I_{c}(0)(1/3)(\surd 3 \xi (0)/r) = 0.8 \ mA(\xi (0)/r)$ at $\xi (0)/r \approx 0.1$. At a radius $r \approx  2 \ \mu m$ of rings used in the measurements, this means that the theoretical amplitude of the persistent current and the amplitude obtained from measurements of critical current oscillations are in agreement if the correlation length $\xi (T) = \xi (0)(1 - T/T_{c})^{-1/2}$ extrapolated to the zero temperature is $\xi (0) \approx  0.2 \ \mu m$. This value does not contradict that expected for the aluminium film structures used in this study, $\xi (0) \approx  0.13 \ \mu m$ in a structure with a smaller ratio of resistances \cite{Mosh1992} and $\xi (0) \approx  1.6 \ \mu m$ for pure aluminium.

Thus, we can conclude that the expected and observed amplitudes of critical current oscillations and their temperature dependences are in agreement, even while oscillations are qualitatively different.

\subsection{Qualitative Difference between Expected and Observed Oscillations of the Critical Current of Asymmetric Superconducting Rings}
The main difference of the observed dependencies $I_{c+}(\Phi /\Phi _{0})$ and $I_{c-}(\Phi /\Phi _{0})$ (Figs. 5, 6, and 8) from expected ones (2) is the similarity of the dependencies of the critical current measured in opposite directions and the positions of extrema (Fig. 8). It follows from the condition of the quantization of the velocity of superconducting pairs (see (1)) that the critical current, independently of the measurement direction and ring asymmetry, should have maximum and minimum values at the total flux inside the ring $\Phi  = n\Phi _{0}$ and $\Phi  = (n+0.5)\Phi _{0}$, respectively. The total magnetic flux $\Phi  = \Phi _{ext} + \Phi _{I}$ is induced by the external magnetic field $\Phi _{ext} = B_{ext}S$ (in our measurements, $B_{ext} = B_{sol} + B_{res}$, see Section 2) and currents $I_{w}$, $I_{n}$ flowing in ring halves. $\Phi _{I} = L_{n}I_{n} - L_{w}I_{w}$, $L_{w} + L_{n} = L$ is the ring inductance, and $L_{w} \approx  L_{n} \approx  L/2$ are the inductances of halves. According to the quantization condition (1), the currents in halves are given by
$$I_{n} = \frac{I_{ext}s_{n}}{s_{n}+s_{w}} + I_{p} = \frac{I_{ext}}{3} + I_{p} $$
$$I_{w} = \frac{I_{ext}s_{w}}{s_{n}+s_{w}} - I_{p} = \frac{2I_{ext}}{3} - I_{p} $$
In a symmetric ring with $s_{w} = s_{n}$, an additional flux is induced only by the persistent current $\Phi _{Ip} = LI_{p}$; in an asymmetric ring with $s_{w} = 2s_{n}$, it is also induced by the external current
$$\Phi _{Iext} = \frac{L_{n}I_{ext}s_{n}}{s_{n}+s_{w}} - \frac{L_{w}I_{ext}s_{w}}{s_{n}+s_{w}} = -\frac{LI_{ext}}{6} $$
$$\Phi _{I} = L_{n}I_{n} - L_{w}I_{w} = \Phi _{Iext} + \Phi _{Ip}$$
We recall that the left-to-right direction corresponds to positive values of $I_{ext}$, $I_{w}$, and $I_{n}$; the clockwise direction corresponds to $I_{p}$. Therefore, for the geometry shown in Fig. 1 (used in this study), $\Phi _{Iext}$ and $\Phi _{Ip}$ have the same sign, while $I_{ext}$ and $I_{p}$ have opposite signs.

The existence of the additional flux $\Phi _{I}$ can shift the positions of extrema by $\Phi _{I}$ in the dependencies on the external magnetic field. The persistent current does not affect the positions of extrema of critical current oscillations $I_{c}(\Phi _{ext}/\Phi _{0})$ in symmetric rings. Numerous studies of symmetric superconducting quantum interferometers, i.e., the same superconducting loops, but with two Josephson junctions being a basis of the DC SQUID (Superconducting Quantum Interference Device) \cite{Barone}, suggest that critical current maxima and minima are observed at $\Phi  = \Phi _{0}$ and $\Phi  = (n+0.5)\Phi _{0}$, respectively. The same is observed for a symmetric aluminium ring without Josephson junctions \cite{Mosh2005}. Asymmetric rings were not studied before our experiments; however, there are papers on oscillations of the critical current of asymmetric interferometers \cite{Barone,Shift}. The difference of the positions of extrema in the magnetic dependences of the critical current from $\Phi  = n\Phi _{0}$ and $\Phi  = (n+0.5)\Phi _{0}$ observed in these studies is associated with the difference of the total flux $\Phi  = \Phi _{ext} +\Phi _{I}$  in the interferometer from the measured $\Phi _{ext}$. Due to interferometer asymmetry, the measuring current $I_{ext}$ induces the additional flux $\Phi _{Iext}$, whose sign varies with the $I_{ext}$ sign. Therefore, the extrema of $I_{c+}(\Phi /\Phi _{0})$ and $I_{c-}(\Phi /\Phi _{0})$ are observed at $\Phi _{ext} = \Phi  - |\Phi _{Iext} | = n\Phi _{0} -  |\Phi _{Iext} |$, $\Phi _{ext} = (n+0.5)\Phi _{0} -  |\Phi _{Iext} |$ and $\Phi _{ext} = n\Phi _{0} +  |\Phi _{Iext} |$, $\Phi _{ext} = (n+0.5)\Phi _{0} +  |\Phi _{Iext} |$, respectively; i.e., the extrema of the dependences of the critical current measured in opposite directions are shifted with respect to each other by $2|\Phi _{Iext} |$.

However, this explanation is inapplicable to the shift  $\Delta \phi$  observed in this study. We obtained the dependencies $I_{c+}(\Phi /\Phi _{0})$ and $I_{c-}(\Phi /\Phi _{0})$ at various temperatures for critical currents $I_{c}$ from $3 \ \mu A$ (Fig. 5) to $30 \ \mu A$ (Fig. 18), which corresponds to the values of $|\Phi _{Iext} | = LI_{c}/6$, differing by an order of magnitude. If the observed shift   $\Delta \phi$ would be determined by the value of $2\Phi _{Iext}$, as is the case in \cite{Shift}, its value should change by an order of magnitude as $I_{c}$ changes from $3 \ \mu A$ to $30 \ \mu A$. However, all our measurements showed that  $\Delta \phi$ is independent of $I_{c}$ (Fig. 18). At the inductance $L = 1.7 \ 10^{-11} \ H$ of the ring used in this study, the additional flux is $\Phi _{Iext} = LI_{ext}/6 \approx  0.04\Phi _{0}$ at $I_{ext} = 30 \ \mu A$ and $\Phi _{Iext} \approx   0.004\Phi _{0}$ at $I_{ext} = 3 \ \mu A$. This estimate confirms the impossibility to explain the observed shift by the difference of the total flux $\Phi $ from the measured flux $\Phi _{ext}$ . The estimates suggest that the used approximation $\Phi  \approx  \Phi _{ext} = B_{ext}S$ is valid with an accuracy of a few percent of $\Phi _{0}$. The magnetic flux $\Phi _{Ip} = LI_{p}$ induced by the persistent current with maximum amplitude $I_{p} = 7 \ \mu A$ (Fig. 17) does not exceed $0.06\Phi _{0}$.

The impossibility of explaining the observed shift by the difference of the total flux in the ring from the measured flux makes it an extraordinarily mysterious phenomenon. The periodicity of the dependencies $I_{c+}(\Phi /\Phi _{0})$ and $I_{c-}(\Phi /\Phi _{0})$ leaves no doubt that the decrease in the critical current is associated with the velocity quantization (1). According to the quantization condition (1), the equilibrium (corresponding to the energy minimum) velocity of superconducting pairs has zero and maximum values at $\Phi  = n\Phi _{0}$ and $\Phi  = (n+0.5)\Phi _{0}$, respectively, in both symmetric and asymmetric rings. Therefore, it seems impossible to explain the shift of extrema by $\pm \Delta \phi /2$ as asymmetry appears in the ring. The results of our measurements of Little-Parks oscillations  $\Delta R(\Phi /\Phi _{0})$ in the studied rings make the situation quite strange. Measurements showed the resistance extrema  $\Delta R(\Phi /\Phi _{0}) \propto v_{s}^{2}(\Phi /\Phi _{0})$ of both the asymmetric and symmetric rings are observed at $\Phi  = n\Phi _{0}$ and $\Phi  = (n+0.5)\Phi _{0}$ (Fig. 18). According to the conventional explanation \cite{Tinkham}, which relates resistance oscillations and the squared velocity of pairs,  $\Delta R(\Phi /\Phi _{0}) \propto v_{s}^{2}(\Phi /\Phi _{0})$, this means that is zero at $\Phi  = n\Phi _{0}$ and is maximum at $\Phi  = (n+0.5)\Phi _{0}$. However, if critical current oscillations $I_{c+}(\Phi /\Phi _{0})$ and $I_{c-}(\Phi /\Phi _{0})$ are also related to pair velocity  $|v_{s} |$ oscillations, their maxima and minima should be observed at $\Phi  = n\Phi _{0}$ and $\Phi  = (n+0.5)\Phi _{0}$, respectively, as in the symmetric ring. For the observed similarity of the dependencies $I_{c-}(\Phi /\Phi _{0}) = I_{c+}(\Phi /\Phi _{0}+\Delta \phi )$, this would imply the absence of CVC asymmetry, $I_{c-}(\Phi /\Phi _{0}) = I_{c+}(\Phi /\Phi _{0})$ at  $\Delta \phi =0$, and the absence of rectified voltage oscillations $V_{dc}(\Phi /\Phi _{0})$. Thus, it should be concluded that the results of measurements of resistance and critical current oscillations on the same asymmetric ring contradict each other based on conventional concepts.

\begin{figure}
\includegraphics{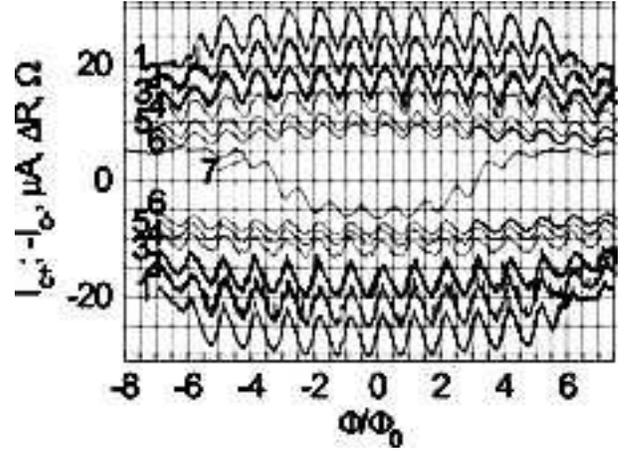}
\caption{\label{fig:epsart} Oscillations of the critical current $I_{c+}(\Phi /\Phi _{0})$, $I_{c-}(\Phi /\Phi _{0})$ and resistance $R(\Phi /\Phi _{0})$ measured on the single ring at various temperatures: (1) $T \approx  1.211 \ K \approx  0.972T_{c}$, (2) $T \approx  1.216 \ K \approx  0.976T_{c}$, (3) $T \approx  1.222 \ K \approx  0.981T_{c}$, (4) $T \approx  1.227 \ K \approx  0.984T_{c}$, (5) $T \approx  1.231 \ K \approx  0.988T_{c}$, (6) $T \approx  1.235 \ K \approx  0.991T_{c}$; (7) $R(\Phi /\Phi _{0})$ measured at $I_{ext} = 1 \ \mu A$ and $T = 1.232 \ K \approx  0.999T_{c}$. The dependence $R(\Phi /\Phi _{0})$ is shifted vertically.}
\end{figure}

\begin{figure}
\includegraphics{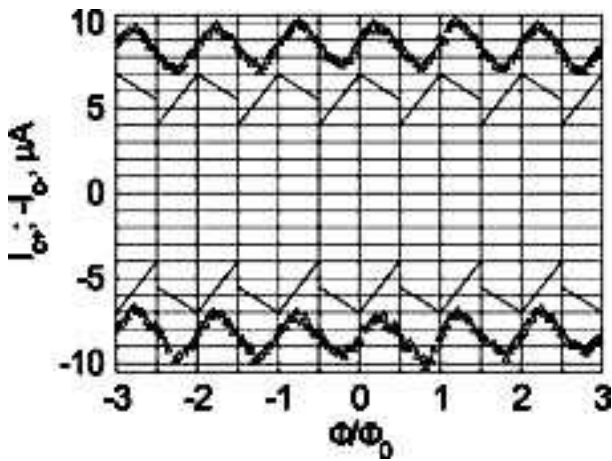}
\caption{\label{fig:epsart} Comparison of the dependences of the critical current: the expected from (2) at $I_{c}(0) = 7 \ \mu A$, $I_{p,A} = 1 \ \mu A$ and the observed on the single ring at $T \approx  1.225 \ K \approx  0.991T_{c}$.}
\end{figure}

The critical current anisotropy in the expected dependencies $I_{c+}(\Phi /\Phi _{0})$ and $I_{c-}(\Phi /\Phi _{0})$ (2) is caused by the absence of similarity (Fig. 19), rather than by their relative shift. In the intervals $\Phi  = n\Phi _{0} \div \Phi  = (n+0.5)\Phi _{0}$ and $\Phi  = (n+0.5)\Phi _{0} \div \Phi  = n\Phi _{0}$, these dependences have different slopes due to the difference between the cross sections of ring halves, $s_{w} = 2s_{n}$; at $\Phi  = (n+0.5)\Phi _{0}$, there is a discontinuity due to a change in the persistent current direction (Fig. 19). The absence of such a discontinuity in the observed dependencies $I_{c+}(\Phi /\Phi _{0})$ and $I_{c-}(\Phi /\Phi _{0})$ (Fig. 19) is one more important difference of the obtained result from the expected one. According to expressions (2a) and (2b), which are based on conventional views, the critical current of the asymmetric ring should depend not only on the magnitude, but also on the direction of the persistent current. According to velocity quantization condition (1), two allowed states with minimum energy, $n$ and $n + 1$, should exist at $\Phi  = (n+0.5)\Phi _{0}$. States with equal in magnitude but oppositely directed persistent currents $I_{p} \propto n - \Phi  /\Phi _{0} = \pm 0.5$ should correspond to different values of the quantum number n. A measurement of the critical current corresponds to a single measurement of the quantum state. When the ring transits to the superconducting state at $\Phi  = (n+0.5)\Phi _{0}$, the quantum number takes the value of either $n$ or $n + 1$, which cannot change before the ring transition to the normal state at $|I_{ext} | = I_{c+}$ or $|I_{ext} | = I_{c-}$. Since the states $n$ and $n + 1$ have opposite directions of the persistent current $I_{p}$, they should correspond to different critical currents,
$$I_{c+} = I_{c0} - I_{p,A}(1 + \frac{s_{w}}{s_{n}}) = I_{c0} - 3I_{p,A} $$
$$I_{c+} = I_{c0} - I_{p,A}(1 + \frac{s_{n}}{s_{w}}) = I_{c0} - 1.5I_{p,A} $$
of the rings under study with the ratio $s_{w}/s_{n} = 2$ (Fig. 19). Since all states, except for $I_{p} = I_{p,A}(n - \Phi /\Phi _{0}) = \pm 0.5I_{p,A}$, are forbidden, the dependences $I_{c+}(\Phi /\Phi _{0})$ and $I_{c-}(\Phi /\Phi _{0})$ should contain a discontinuity whose value $1.5I_{p, A}$ should be equal (according to (2)) to half the amplitude of the critical current oscillations (Fig. 19). However, we observe oscillations $I_{c+}(\Phi /\Phi _{0})$ and $I_{c+}(\Phi /\Phi _{0})$ (Figs. 6, 8, and 18) which, in contrast to the expected ones (2), contain no discontinuity at $\Phi  = (n+0.5)\Phi _{0}$ (Fig. 19).

This result of a single measurement is even stranger, since multiple measurements show two allowed states at $\Phi  = (n+0.5)\Phi _{0}$. Here it is noteworthy that the dependencies of the rectified voltage $V_{dc}(\Phi /\Phi _{0})$ cross zero at $\Phi  = n\Phi _{0}$ and $\Phi  = (n+0.5)\Phi _{0}$ (Figs. 9-11), while resistance oscillations  $\Delta R(\Phi /\Phi _{0})$ have a minimum and a maximum at $\Phi  = n\Phi _{0}$ and $\Phi  = (n+0.5)\Phi _{0}$, respectively (Fig. 18). The rectified voltage is proportional to the multiply measured pair velocity, $V_{dc}(\Phi /\Phi _{0})  \propto  \overline{v_{s}} \propto \overline{n - \Phi /\Phi _{0}}$, while the resistance is proportional to the squared velocity,  $\Delta R(\Phi /\Phi _{0})  \propto  \overline{v_{s}^{2}} \propto \overline{(n - \Phi /\Phi _{0})^{2}}$. The   $\Delta R(\Phi /\Phi _{0})  \propto  \overline{v_{s}^{2}} $ minimum and $V_{dc}(\Phi /\Phi _{0})  \propto  \overline{v_{s}} = 0$ at $\Phi  = n\Phi _{0}$ correspond to the contribution of only one level. However, the $\Delta R(\Phi /\Phi _{0})  \propto  \overline{v_{s}^{2}} $ maximum at $V_{dc}(\Phi /\Phi _{0})  \propto  \overline{v_{s}} = 0$ cannot be explained without the existence of two states $n$ and $n + 1$ at $\Phi  = (n+0.5)\Phi _{0}$ with equal and oppositely directed velocities $ \overline{v_{s}} \propto \overline{n - \Phi /\Phi _{0}} \propto  (1/2) + (-1/2)$, while $\overline{v_{s}^{2}} \propto \overline{(n - \Phi /\Phi _{0})^{2}} \propto (1/2)^{2} + (-1/2)^{2} = 1/2$. The detected disagreement between the results of single and multiple measurements of states of the macroscopic quantum system can be of great importance for the problem of quantum calculations \cite{Valiev}, in particular, the problem of developing the quantum bit on the basis of superconductor nanostructures \cite{Makhlin}.

\section{CONCLUSIONS}
The goal of this study was to test the quite obvious assumptions that (i) quantum oscillations of the rectified voltage $V_{dc}(\Phi /\Phi _{0})$ are caused by periodic variations in the CVC asymmetry of asymmetric superconducting rings in a magnetic field; (ii) the CVC asymmetry is caused by the superposition of the measuring and persistent currents, and the periodic CVC variation is caused by periodic variations of the persistent current, $I_{p}(\Phi /\Phi _{0})$. The results obtained not only confirmed the first assumption, but also allowed us to explain the change in the rectification efficiency with the current amplitude and temperature. As for the second assumption, the periodic dependence of the critical current on the magnetic field leaves no doubt that the CVC asymmetry is caused by quantization of the superconducting pair velocity, i.e., the periodic dependence of the persistent current. Many results confirm the assumption on the CVC asymmetry as a consequence of the superposition of measuring and persistent currents. However, the similarity of the dependencies $I_{c+}(\Phi /\Phi _{0})$ and $I_{c-}(\Phi /\Phi _{0})$, their shift with respect to each other, and the absence of discontinuity at $\Phi  = (n+0.5)\Phi _{0}$ are not only inconsistent with the expected dependence, but also contradict the measurements of Little-Parks resistance oscillations. The cause of this contradiction should be clarified in further studies.

\section*{Acknowledgement}
This study was supported by the Russian Foundation for Basic Research, project no. 04-02-17068, the basic research program of the Department of Information Technologies and Computing Systems of the Russian Academy of Sciences "Organization of Calculations Using New Physical Principles" within the project "Quantum Bit Based on Micro- and Nanostructures with Metallic Conductivity", and the program of the Presidium of the Russian Academy of Sciences "Low- Dimensional Quantum Structures."

\end{document}